\documentclass[amsmath,twocolumn,superscriptaddress]{revtex4-1}
\usepackage{epsfig}
\usepackage{amssymb}
\usepackage{amsmath}
\usepackage{amsfonts}
\usepackage{braket}
\usepackage[T1]{fontenc}
\usepackage{bm}
\usepackage{graphicx}
\usepackage{xcolor}
\usepackage{verbatim}
\usepackage{soul}
\usepackage{tikz}
\usetikzlibrary{matrix,shapes,arrows,positioning,chains}
\usetikzlibrary{shapes.geometric}
\usetikzlibrary{shapes.arrows}
\usepackage{array}
\usepackage{enumitem}
\usepackage{color}
\usepackage{physics}
\usepackage[export]{adjustbox}
\usepackage[percent]{overpic}

\graphicspath{{./}{fergus/}}

\begin{document}
\title{Parallel Quantum Simulation of Large Systems on Small NISQ Computers}

\author{F. Barratt}
\affiliation{Department of Mathematics, King's College London, Strand, London WC2R 2LS, United Kingdom}

\author{James Dborin}
\affiliation{London Centre for Nanotechnology, University College London, Gordon St., London, WC1H 0AH, United Kingdom}

\author{Matthias Bal}
\affiliation{GTN Limited, Clifton House, 46 Clifton Terrace, Finsbury Park, London N4 3JP, United Kingdom}

\author{Vid Stojevic}
\affiliation{GTN Limited, Clifton House, 46 Clifton Terrace, Finsbury Park, London N4 3JP, United Kingdom}

\author{Frank Pollmann}
\affiliation{Department of Physics, T42, Technische Universit\"at M\"unchen, James-Franck-Stra{\ss}e 1, D-85748 Garching, Germany}

\author{A.~G. Green}
\affiliation{London Centre for Nanotechnology, University College London, Gordon St., London, WC1H 0AH, United Kingdom}
\affiliation{email: andrew.green@ucl.ac.uk}

\date{\today}
\begin{abstract}
Tensor networks permit computational and entanglement resources to be concentrated in interesting regions of Hilbert space. Implemented on NISQ machines they allow simulation of quantum systems that are much larger than the computational machine itself. This is achieved by parallelising the quantum simulation. 
Here, we demonstrate this in the simplest case; an infinite, translationally invariant quantum spin chain. 
We provide Cirq and Qiskit code that translate infinite, translationally invariant matrix product state (iMPS) algorithms to finite-depth quantum circuit machines, allowing the representation, optimisation and evolution arbitrary one-dimensional systems. Illustrative simulated output of these codes for achievable circuit sizes is given.
\end{abstract}
\maketitle

\vspace{0.2in}

 \section{Introduction} 
 \label{sec:introduction}
The insight underpinning Steve White's formulation of the density matrix renormalisation group (DMRG) is that entanglement is  the correct resource  to focus upon to formulate accurate, approximate descriptions of large quantum systems\cite{SteveWhiteDMRG}. 
Later understood as an algorithm to optimise a matrix product state (MPS)\cite{OstlundRommer}, this notion underpins the use of tensor networks as a variational parametrisation of wavefunctions with quantified entanglement resource. 
Such approaches allow one to concentrate computational resources in the appropriate region of Hilbert space and provide an effective and universal way to simulate quantum systems \cite{mps_representations,mps_classically_efficient}. 
They  also provide an effective framework to distribute entanglement resources in simulation on noisy intermediate scale quantum (NISQ) computers.

\begin{figure}[ht]
\includegraphics[width=0.45\textwidth]{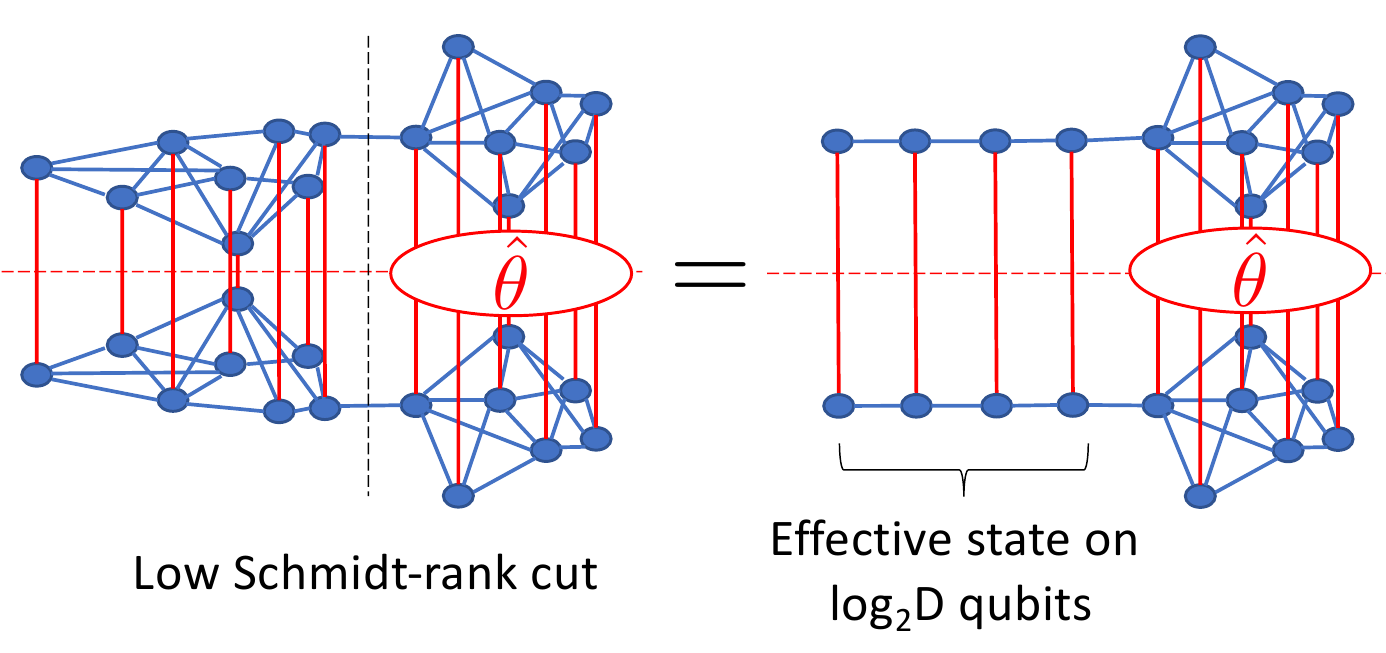}
\caption{{\bf Tensor network for a quantum state that is weakly entangled across a certain partition}. This weak entanglement allows parallel simulation of the two partitions of the system. The expectation of an operator located to the right of the partition can be carried out by replacing the state on the left by a state over much fewer spins (the number determined by the entanglement across the cut). The numerical values of correlations in this smaller representation of the left are determined by quantum effects in the full left hand system, and can be computed in parallel and iterated to consistency.}
\label{fig:ParrallelQSim}
\end{figure}

Quantum computers such as those of Google, Rigetti, IBM and others implement finite-depth quantum circuits with controllable local two-qubit unitary gates. 
Innovations for quantum simulation include using these circuits as variational wavefunctions\cite{VQE}, optimising them stochastically or by phase estimation\cite{phase_estimation}, and evolving them either by accurate Trotterisation of the evolution operator\cite{trotter1959product,SUZUKI1993232,hastings2014improving} or variationally\cite{BenjaminTDVP}. 
Currently available NISQ devices are limited by gate fidelity and the resultant restriction of available entanglement resources. 
Since the finite-depth quantum circuit may be equivalently described as a tensor network\cite{circuits_are_mps}, tensor networks provide a convenient framework with which to distribute entanglement to the useful regions of Hilbert space and to make efficient use of this relatively scarce resource. 

We dub the implementation of a tensor network on such a NISQ device a {\it Quantum tensor network}. 
There are several advantages to this framework. 
It fits directly into a broader ecosystem of classical simulation of quantum systems. 
Indeed, because it is based upon the manipulation of explicitly unitary elements, the quantum circuit provides perhaps the most natural realisation of tensor networks. 
Canonicalisation at each step in a classical tensor network calculation amounts to reducing the tensors to isometries --- a step that is not required in an explicitly unitary realisation. 
Moreover, the remaining elements of unitaries parametrise the tangent space of the variational manifold\cite{haegemanTDVP,mps_path_integrals}.

Here we  demonstrate that quantum tensor networks can be used to parallelise quantum simulation of systems that are much larger than available NISQ machines\cite{peng2020simulating, kim2017holographic, kim2017noise}. 
Central to this is dividing the quantum system into a number of sub-elements that are weakly-entangled and can be simulated in parallel on different circuits. 
The influence of the different regions of the system upon one another can be summarised by an effective state on a much smaller number of quantum bits. 
We provide Cirq and Qiskit code for the simplest class of examples --- infinite, translationally invariant quantum spin chains. 
This is a direct translation ({\it mutatis mutandis}) of iMPS algorithms to quantum circuit machines. 
The remarkably simple circuits revealed below allow the representation of an infinite quantum state, and its optimisation and real-time evolution for a given Hamiltonian. 

\section{Results}

 \subsection{Parallel Quantum Simulation Across Weakly-Entangled Cuts}
 \label{sec:ParallelQC}
To parallelise our simulation on a small NISQ machine, we first identify partitions of the system where the effect of one partition upon the other can be summarised by a small amount of information. 
This is achieved by making Schmidt decompositions across the cut:
$| \psi \rangle = \sum_{\alpha=1}^D \lambda^\alpha | \phi^\alpha_L \rangle  | \phi^\alpha_R \rangle, $
where $ | \phi^\alpha_L \rangle$ are an orthonormal set of states to the left of the cut and $ | \phi^\alpha_R \rangle$ the same on the right. 
The $\lambda^\alpha$ are known as the Schmidt coefficients and $D$ the Schmidt rank or bond order. 
Retaining $\lambda^\alpha$ only above some threshold value provides a way to compress representations of a quantum state; the MPS construction can be obtained by applying this procedure sequentially along a spin chain\cite{mps_classically_efficient}. 

If an observation is made on the right-hand-side of such a cut, the effect of the quantum state on the left upon the observation can be summarised by just $D$ variables corresponding to the Schmidt coefficients. 
This same effect can be achieved by an effective state on a spin chain of length $\log_2 D$  --- see Fig. \ref{fig:ParrallelQSim}  ---which can be parametrised on the quantum circuit by an $SU(D^2)$ unitary $V_L$. 
This encodes both the Schmidt coefficients $\lambda_\alpha$ and the orthonormal states $|\phi^\alpha_L \rangle$. The latter do not contribute to observables on the right and so in principle $V_L$ can be parametrised by just $D$ variational parameters. The precise numerical values must be determined by solving a  quantum mechanical problem on the left of the system.  Similarly for observations made to the left of the cut, the effect of the right-hand side can be summarised by a unitary $V_R$. 

This gives a prescription for parallel quantum simulation. Calculations of the quantum wavefunction to the left and right of the cut can be carried out on different  quantum circuits or sequentially on the same circuit. 
The effects of the left partition upon the right partition and {\it vice versa} --- through the environment unitaries $V_L$ and $V_R$ --- are iterated to consistency. At each stage of this iteration, measurements must be performed in order to determine $V_{L/R}$. 
The small Schmidt rank of the cut reduces the computational complexity of this process - if we were to do full state tomography, to $\mathcal{O}(\mathrm{poly}(D))$, but with more sophisticated methods even to $\mathcal{O}(\log(D))$, as in the example in the following section.

There are many physical situations in which this parallelisation might be useful. For example, large organic molecules that have localised chemical activity --- this activity may be modulated or tuned by the surrounding parts of the molecule and the interplay of these effects could be calculated in parallel.
In the following, inspired by iMPS tensor networks, we give quantum circuits that embody these ideas.

\subsection{Parallel Simulation with Quantum Tensor Networks}

 \begin{figure}
     \centering
   \includegraphics[width=0.85\linewidth]{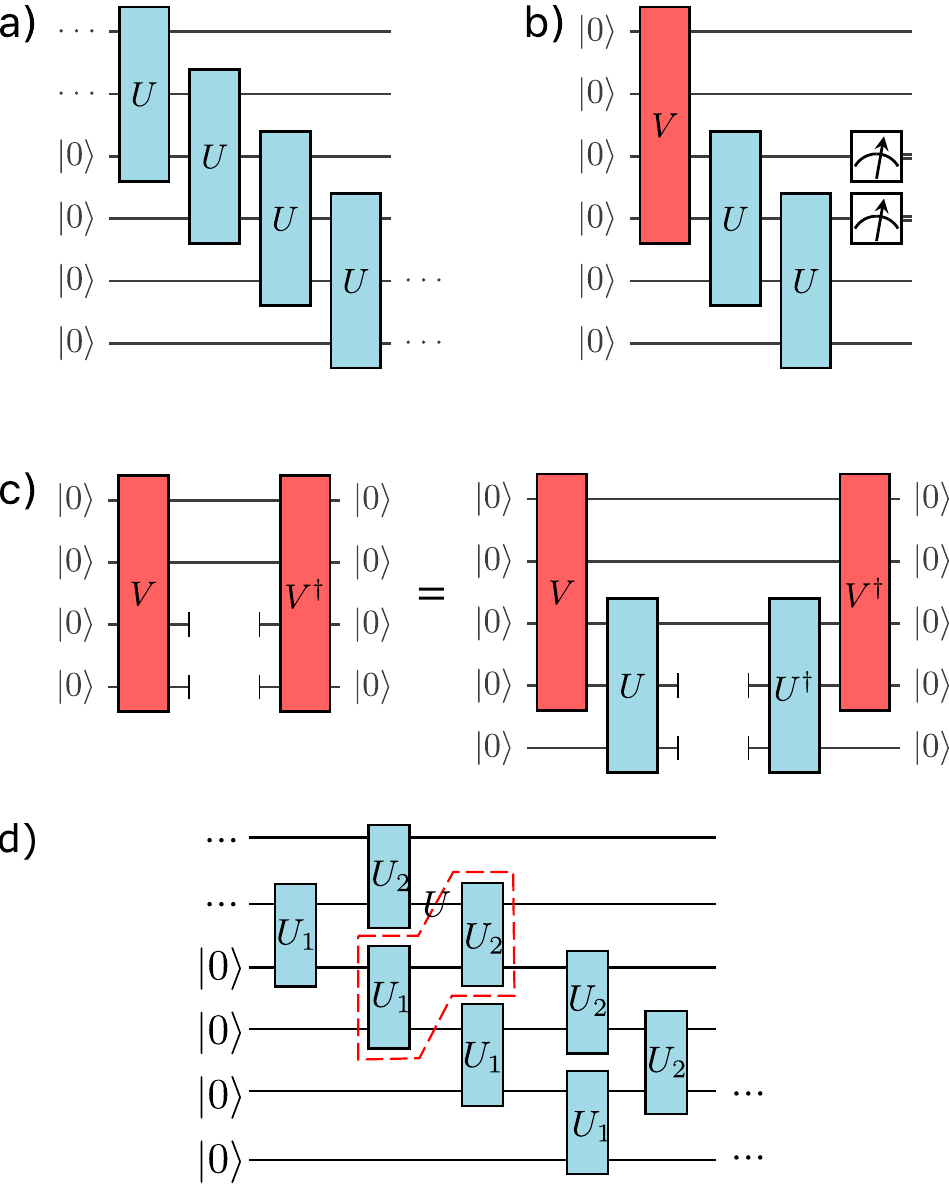}
\caption{{\bf Quantum circuits for translationally invariant states and their local measurement.} a) An infinite depth and width quantum circuit representing a translationally invariant state. $U \in SU(dD)$ with $d$ the local Hilbert space dimension and $D=2^N$ the bond order. $d=2$ for spin $1/2$ and is used exclusively throughout this paper. In these illustrations $D=4$. The circuit acts upon a reference state $|000...\rangle$ at the left of the figure with unitary operators applied sequentially reading left to right. b) Local measurements on this translationally invariant state can be reproduce exactly by the finite circuit shown. The reduced form takes advantage of the unitarity of $U$, due to which sites to the left of the observable do not contribute,. The environment unitary $V\equiv V(U) \in SU(D^2)$ summarises the effect of sites to the right of the observable and describes an effective state over $N=\log_2 D$ spins. c) The environment unitary $V(U)$ is the solution of the fixed point equation shown. This equation is to be interpreted as an equality of the reduced density matrices implied by the free qubit lines. We show in the Methods how to implement this using swap gates. d) A shallow circuit representation of the $D=4$ state following Ref.[\onlinecite{lin2021real}]. Such circuits have been shown to have high fidelity with states obtained in Hamiltonian evolution and are exponentially quicker to contract on a quantum circuit than they are classically.}
\label{fig:QiMPS}
\end{figure}

 \begin{figure*}
     \includegraphics[width=1\linewidth]{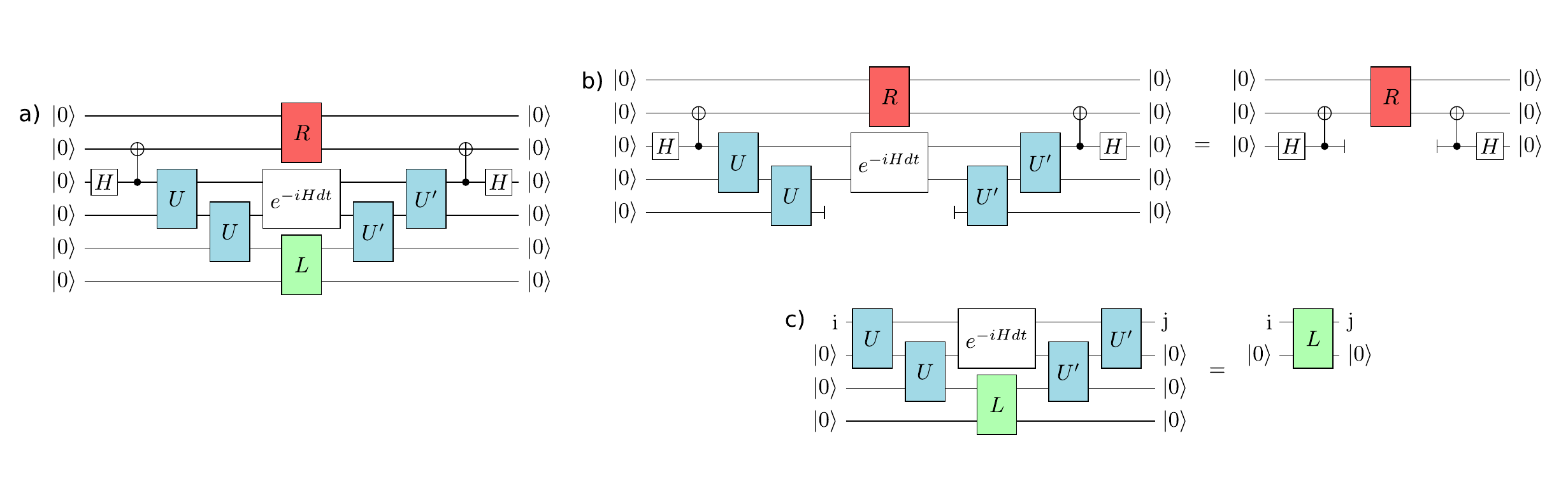}
\caption{{\bf Quantum Implementation of the Time-dependent variational principle.} For simplicity, we depict the above circuits for $D=2$. Higher bond order cases are given in the supplementary materials. a) The unitary $U'$ that optimises the overlap of this circuit with $|000... \rangle$ describes the time evolution of the state described by $U(t)$ by a time interval $dt$ under the Hamiltonian ${\cal H}$, {\it i.e.} $U'=U(t+dt)$. b) and c) The mixed environment unitaries $R$ and $L$ are given by the fixed point solutions of these circuit equations. As in Fig.\ref{fig:QiMPS}, these are to be interpreted as an equality of the density matrices implied by free qubit lines.}
\label{fig:QTDVP}
\end{figure*}

The translationally invariant MPS gives an approximate representation of a translationally invariant spin {1/2} state as 
\begin{equation}
|\psi \rangle = \sum_{ \{i \}, \{ \sigma \}} \prod_n ... \;
A^{\sigma_{n-1}}_{i_{n-1},i_{n}}  A^{\sigma_n}_{i_n,i_{n+1}} 
... 
| ... \sigma_{n-1},\sigma_n ... \rangle,
\label{eq:MPSstate}
\end{equation}
where $n$ labels the lattice site, $\sigma_n$ labels the spin {1/2} basis states on the $n^{th}$ site and $i_n$ are auxiliary tensor indices that run from $1$ to the bond order $D$. The tensors $A^\sigma_{ij}$ are the same on every site reflecting the translational invariance. These states can be mapped to quantum circuits by taking the MPS tensors in isometric form and identifying them with the circuit unitaries as described in the Methods. 
A variety of techniques have been developed for the classical manipulation of these states for quantum simulation\cite{mps_representations,schollwock_review,orus_review}. Here we confine ourselves to discussing the quantum circuit realisation. The Supplementary Material gives a detailed summary of the connection between the classical MPSs and their quantum circuit realisation.

 \noindent
{\it Representing the state}:
 A translationally invariant, spin {1/2} MPS state of bond order $D=2^N$ can be represented by the infinite circuit shown in Fig.\ref{fig:QiMPS} a).  Expectations of local operators in this state can be evaluated by the finite circuit shown in Fig.\ref{fig:QiMPS} b). The effect of contracting the infinite circuit to the left of the operator is trivial due to the the unitarity of $U\in SU(2D)$ (which automatically encodes the left canonical form of the related MPS tensor). The contraction to the right is described by the tensor $V \in SU(D^2)$, which encodes an effective state over $N=\log_2 D$ spins and their entanglement with the remaining system to the left-hand side. This unitary is determined self-consistently from $U$ by the circuit shown in Fig. \ref{fig:QiMPS} c). 
 As demonstrated in Ref.\cite{mps_classically_efficient}, an MPS in this form can be constructed from any state by a sequence of Schmidt decompositions running from left to right. This guarantees the existence of the isometric MPS representation and the quantum circuit realisation of it.
 The operation of such a circuit at $D=2$ was demonstrated in Ref.\cite{Smith2019} on an IBM quantum circuit, where analytic forms where known for both $U$ and $V$ along a line through the phase diagram of a model with topological phase transition. In general, $V \equiv V(U)$ is not known and must be solved following Fig. \ref{fig:QiMPS} c).

\vspace{0.1in}
 \noindent
{\it Optimising the state}:
We can find the ground state and the corresponding energy density of translationally invariant Hamiltonians by minimizing the expectation value of the energy. The algorithm mirrors the variational quantum eigensolver. 
The expectation of the local Hamiltonian is found by measuring the corresponding Pauli strings on the physical qubits (see Fig.~\ref{fig:QiMPS}b)). The result can then be minimised as a function of the ansatz parameters.
Updates must be interleaved with updates to the environment, $V$, such that we optimize over valid translationally invariant states.

\vspace{0.1in}
 \noindent
{\it Evolving the state}:
Perhaps the most compelling feature of this implementation is the ease with which time-evolution can be achieved. The simple circuit shown in Fig. \ref{fig:QTDVP} a) returns the unitary $U'\equiv U(t+dt)$ that updates the state encoded by $U(t)$ to a time $t+dt$ under evolution with the Hamiltonian ${\cal H}$. The first variation of this circuit with respect to $U'$ returns the time-dependent variational principle for iMPS in the form first presented by Haegeman {\it et al.} in Ref. \cite{haegemanTDVP}. The equivalence uses the automatic encoding of the gauge-fixing of the state to canonical form as well as encoding of the tangent space and its gauge fixing (see Methods section and additional notes in Supplementary Materials). As in the determination of the best groundstate approximation above, the update involves two nested loops; one to find the update $U'$ and one to find the environment tensors $L\equiv L(U,U')$ and $R \equiv R(U,U')$ --- both of which are required in this case as the circuit corresponds to the overlap of two different states rather than expectations taken in a given state. We have used a slightly different way of representing these environments in Fig. \ref{fig:QTDVP} compared to that employed in Fig. \ref{fig:QiMPS}.

 \vspace{0.1in}
 \noindent

{\it Quantum Advantage}:
It is natural to ask whether there is any quantum advantage from using a quantum circuit in this way.  Algorithms for manipulating iMPS (iDMRG, TDVP, etc.) \cite{orus_review, mps_representations, schollwock_review} are classically efficient - they have complexity of $\mathcal{O}(D^3)$. 
Where then is the room for improvement by implementation on a quantum circuit?
The quantum advantage comes from the potentially exponential reduction in the dependence upon the bond dimension, $D$.

In a quantum circuit, the multi-qubit unitaries must be compiled to the available gate set. A translationally invariant state with entanglement $S$ can captured by a matrix product state of bond dimension $D \sim \exp S$. This requires a circuit depth of $\mathcal{O}( \exp S)$.
An arbitrary $U \in SU(2 \exp S)$ unitary to implement this iMPS, requires $\mathcal{O}(\exp S)\sim \mathcal{O}(D)$ gates \cite{nielsen_chuang}.  However, a subset of non-trivial unitaries with entanglement $S$ can be achieved with circuits whose depth is $\mathcal{O} (S) \sim \mathcal{O}(\log D)$  giving an exponential speedup over the classical implementation\cite{cerezo2021cost}. 
This reduces the contraction time from $\mathcal{O}(D^3)$ for a typical classical implementation to $\mathcal{O}(\log D)$ in the quantum case. Though these shallow circuits exist, the question remains whether they have high fidelity with the states that we are interested in. Ref.[\onlinecite{lin2021real}] identifies a subset of shallow quantum circuit MPS that have high fidelity with the states produced by Hamiltonian evolution. An example at bond order $D=4$ is shown in Fig. \ref{fig:QiMPS} d). 

This demonstrates a quantum advantage for $U$. We must also consider whether the environment $V$ (or $R$ and $L$) have high fidelity shallow circuit representations. {\it A priori} there is no guarantee that, given a shallow circuit $U$, the $V$ that satisfies the fixed point equations in Fig.~\ref{fig:QiMPS}, or the $R$ and $L$ that satisfy the fixed point equations in Fig.~\ref{fig:QTDVP} are themselves shallow . However, there is a rigorous argument for this. We start by constructing an initial shallow approximation. When optimising the energy, as in Fig.~\ref{fig:QiMPS}, we can take advantage of the ability to diagonalize the reduced density matrix to the right and choose a corresponding $V$. A shallow circuit approximation to this allows us to set $\log D$ Schmidt coefficients exactly with the remainder lying on a smooth interpolation between them. 
In the case of time evolution, the mixed environments  $R$ and $L$ cannot necessarily be diagonalised simultaneously (though off-diagonal elements are of order $dt^2$) and a richer --- though still shallow --- variational parametrization allowing for this is necessary. In either case, a shallow approximation for the environment can be improved exponentially for a linear cost in qubits and circuit depth by applying the transfer matrix a linear number of times, {\it i.e.} by using the power method (see the Supplementary Materials for more details). This simply corresponds to inserting further copies of the transfer matrix in the centre of the circuits in Fig.~\ref{fig:QTDVP} a.
These arguments establish an asymptotic quantum advantage for our algorithm. In practice, we find that the initial shallow approximations prove remarkably accurate and these corrections are unnecessary. 

\subsection{Numerical Results}
 \label{sec:Results}

 \begin{figure}
     \centering
     \includegraphics[width=\linewidth]{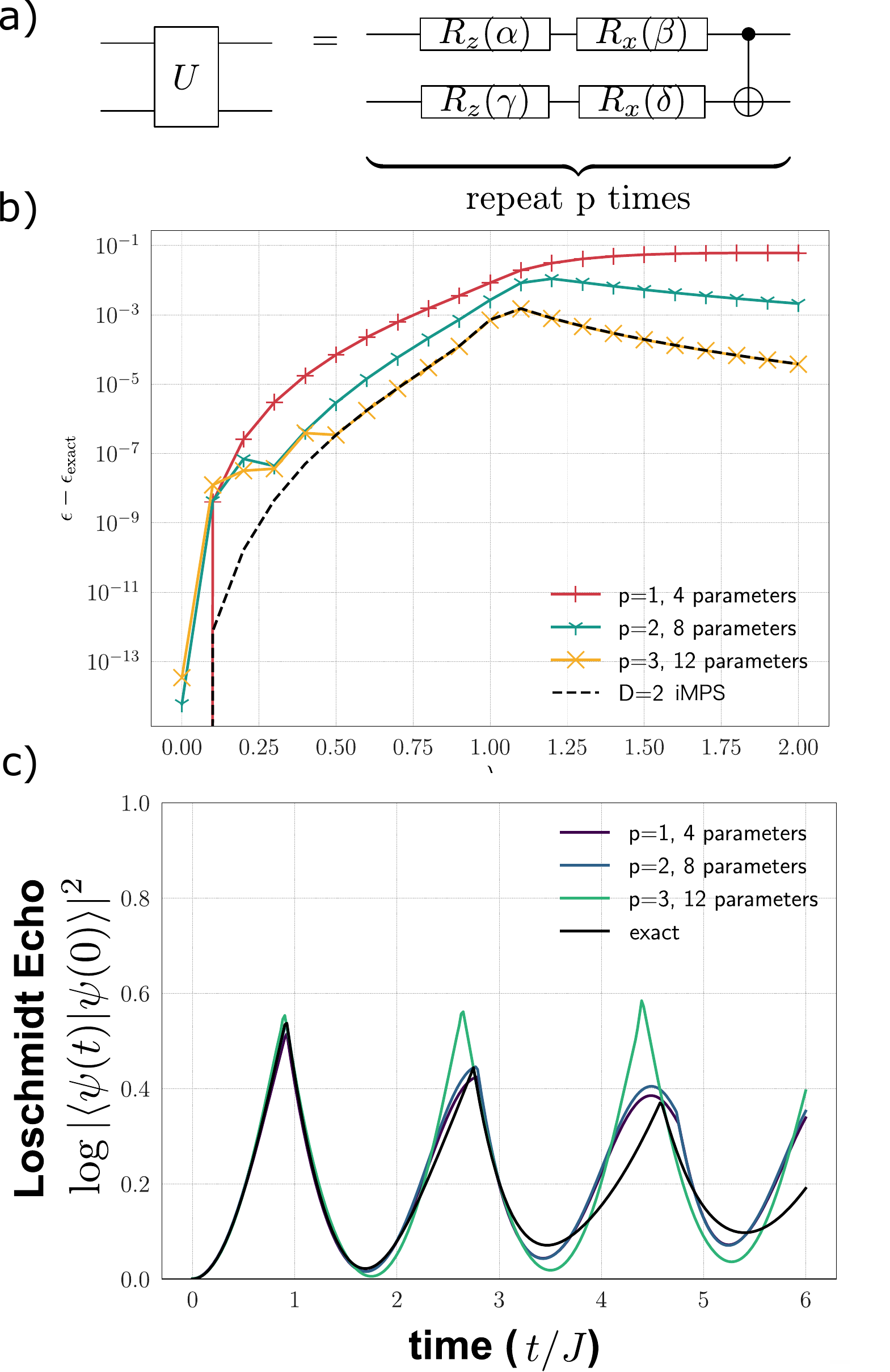}
\caption{{\bf Results of simulating the transverse field Ising model:}  
The Hamiltonian ${\cal H}= \sum_n \left[ \hat \sigma^z_{n}  \hat \sigma^z_{n+1} + \lambda \hat \sigma^x_n \right]$ is studied with a bond order $D=2$ quantum matrix product state. a) The $SU(4)$ unitaries $U$ and $V$ are compiled to the circuit as shown. The parameter $p$ is varied to increase the accuracy. 
Although more efficient parametrizations exist for 2 qubit unitaries\cite{vatan2004optimal}, as well as circuits more specifically tuned to this problem, we choose a generic circuit. It is readily extendible to higher bond orders [see Supplementary Materials].
 b) The optimum state is found using the circuits depicted in Fig. \ref{fig:QiMPS}. The energy of this state is a better approximation to the true groundstate energy as the depth of parametrization increases and converges to that obtained in a conventional MPS algorithm. In particular, we have checked that the parametrization of Ref.\cite{vatan2004optimal} perfectly reproduces the MPS results. Note that at $\lambda =0$ the Hamiltonian is optimised by a product state, which is captured perfectly with $p=1$. 
c) Transverse field Ising model displays dynamical phase transitions in the Loschmidt echo\cite{dynamical_phase_transitions}. These are revealed in the simulated runs of the quantum time-dependent variational principle embodied by the circuits in Fig.\ref{fig:QTDVP}. More accurate circuits are required to obtain good agreement.
The results indicated as exact in the above are exact analytical results.}
\label{fig:TransverseIsingResults}
\end{figure}

We have written Cirq and Qiskit code to implement the quantum circuits shown in Figs.~\ref{fig:QiMPS} and \ref{fig:QTDVP}. The results of running this code in simulation on Google's Cirq simulator are shown in Fig.~\ref{fig:TransverseIsingResults}. 
We have chosen optimisation and time evolution of the transverse field Ising model\cite{dynamical_phase_transitions}, and Poincare sections of the dynamics of the PXP Hamiltonian\cite{michailidis2020slow} as illustrative examples. 
The properties of the transverse field Ising model are well understood. 
The Loschmidt echo (fidelity of the time-evolved wavefunction with the initial wavefunction)  reveals a dynamical phase transition\cite{dynamical_phase_transitions} which provides a non-trivial test for our simulation. Our main findings are as follows:

\noindent
i. When run without gate errors and complete representation of the unitaries $U$,$V$, $L$, and $R$, our code precisely reproduces the optimum iMPS and its time evolution using the time-dependent variational principle.

\noindent
ii. Factorisations of the unitaries reduce the fidelity of our results. These are systematically improved as the depth or expressibility of the ansatz is increased. 
Full parametrisations of the unitaries reproduce classical MPS results exactly. 

\noindent
iii. Going from representing, to optimising, to time-evolving states places increasing demands upon circuit-depth and width. Accurate results require increasingly deep factorisation of the unitaries, and suffer increasingly adverse effects of gate errors.

\vspace{0.1in}
\noindent
{\it Optimising}: The simulation results for the optimisation of the transverse field Ising model are given in Fig.~\ref{fig:TransverseIsingResults}. The unitary $U$ is factorized using the ansatz in Fig.~\ref{fig:TransverseIsingResults}a. This ansatz shows good agreement with the exact bond dimension 2 MPS results with low depth circuit. This factorisation is much lower depth than a full factorisation of SU(4). Alternative factorisations of the state unitary are possible for this problem. Lower depth ansatz can be used by identifying that the ground state of the transverse Field Ising model is made up of real tensors (See Supplementary Material for details).

\vspace{0.1in}
\noindent
{\it Time Evolving}: Using exact representations of unitary matrices we exactly reproduce the TDVP equations as expected. We demonstrate that shallow factorisations of the state unitaries are able to effectively capture dynamical phase transitions in the Loschmidt Echos. The ground state of the transverse field Ising model with $g=1$ is prepared. This state is then evolved under the same Hamiltonian but with $g=0.2$. At each step of the trajectory, the overlap of the current state with the original state is recorded. The log of this overlap is shown for the whole trajectory in Fig.~\ref{fig:TransverseIsingResults}c.

\begin{figure}
     \centering
     \includegraphics[width=\linewidth]{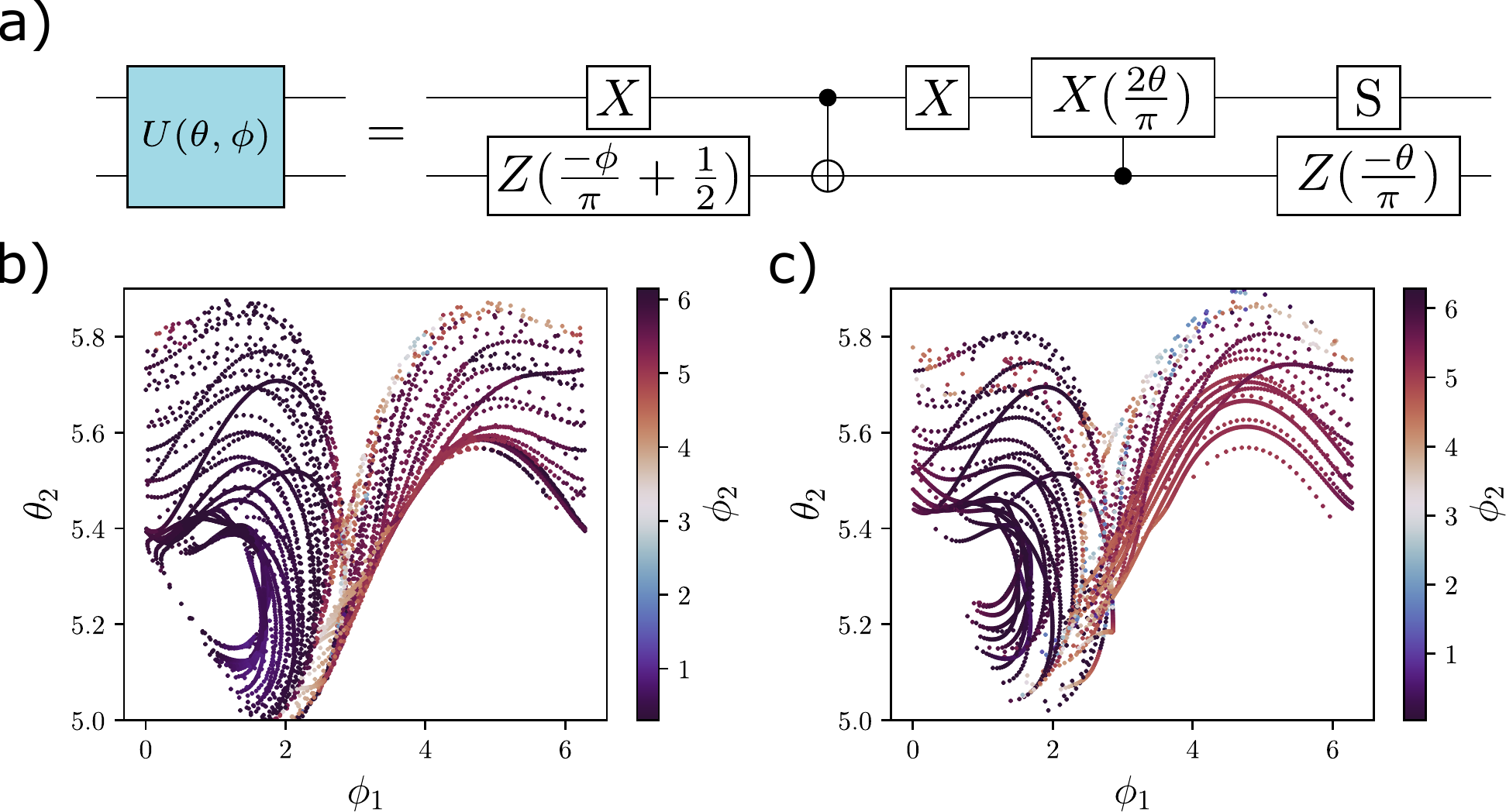}
\caption{{\bf Many Body Scars in the PXP model:}  
The Hamiltonian ${\cal H}= \sum_n (1-\hat \sigma^z_{n-1})\hat \sigma_n^x (1- \hat \sigma^z_{n+1})$, first posited to describe the results of quantum simulations using Rydberg atoms\cite{turner2018weak}, displays a curious property known as many-body scarring\cite{turner2018weak}, whereby from certain starting states, persistent oscillations that can be described with a low bond-order MPS are found. These are amenable to study on a NISQ machine using the quantum time-dependent variational principle of Fig \ref{fig:QTDVP}. a) A simple set of 2-site periodic states at bond order $D=2$ are parametrized by circuits with just 2 parameters per site, so 4 in total. b) A partial Poincare section through the plane $\theta_1=0.9$ produced from a classical simulation using the matrix product state equations of motion presented in Ref.\cite{michailidis2020slow}. Initial conditions are chosen on a constant energy surface $\langle {\cal H}\rangle = 0$. The partial plot was produced with initial conditions along a line with spacing $\delta \phi_1 = 3\times 10^{-2}$, with $\theta_1 = 0.9$ and $\theta_2 = 5.41$. The final parameter, $\phi_3$, chosen to fix the energy. 
 c)The same Poincare map produced simulating the quantum time-dependent variational principle in Cirq. While the figure is blurred somewhat by integration, the main features are still apparent}
\label{fig:ScarResults}
\end{figure}

A low depth state is able to exactly represent the states used in Ref.~\cite{michailidis2020slow} to produce Poincare maps, which are used to study slow quantum thermalisation in the PXP Model\cite{turner2018weak}. The states are defined with a two parameter circuit shown in Fig.~\ref{fig:ScarResults}. The quantum TDVP code is able to recreate the Poincare maps for a two site translationally invariant MPS state. It is possible in this case to discard erroneous points by identifying points with energies that deviate from the known value by more than some fixed threshold. This is a form of error mitigation which may be applicable to other problems when using the quantum TDVP algorithm, and may help mitigate the impact of noise from NISQ devices. The larger structures in the Poincare Map are distorted by errors, but are still visible in the presence of integration and stochastic optimisation errors. The effects of noise on the quantum TDVP algorithm are outlined in the Supplementary Materials.

\section{Discussion}
 \label{sec:Discussion}

We have presented a way to perform quantum simulations by translating tensor network algorithms to quantum circuits. Our approach allows parallel quantum simulations of large systems on small NISQ computers. We have demonstrated this for one-dimensional translationally invariant spin chains. The translation of MPS algorithms naturally encodes fundamental features of matrix product states and the tangent space to the variational manifold that they form. In demonstrating the operation of such circuits, we have touched upon some immediate questions including the expressiveness of shallow circuit restrictions of tensor network states, their effect upon simulation alongside that of  finite gate fidelity. These warrant further systematic study.  

Our algorithms are readily extensible to inhomogeneous one-dimensional systems and to higher dimensions following existing methods that wrap one-dimensional states to higher dimensional systems\cite{schollwock_review,orus_review}. It would be interesting to study other gauge restrictions of MPS --- such as the mixed gauge of modern classical time-dependent variational principle codes --- which can also be implemented in quantum circuits. Generalisations of MPS that more directly describe higher dimensional systems are also available. For example, the projected entangled pair states (PEPS) give a two-dimensional generalisation. In realising these states on a quantum circuit, they must be formed from isometric tensors. Until recently, a suitable canonical form for PEPS was not available. The isometric version of PEPS presented in Ref.\cite{zaletel2020isometric} shows great promise and ought to be possible to implement on a suitably connected quantum circuit. Other tensor networks such as the multi-scale entanglement renormalisation ansatz \cite{mera_vidal} (MERA) are naturally based upon unitary operators and can be realised on a quantum circuit\cite{kim2017robust}. Indeed, MERA has been deployed for image classification on a small quantum circuit\cite{hierarchical_quantum_classifiers} and as a quantum convolutional neural network\cite{cong2019quantum}.

The tensor network framework also provides a convenient route to harness potential quantum advantage in simulation. The one-dimensional matrix product state ansatz is efficiently contractible. The time taken to calculate the expectation of a local operator scales proportional to the length of the system. A quantum implementation has the advantage of a potentially exponential decrease in the prefactor to this scaling. While a classical tensor network may efficiently represent the important correlations of quantum state in higher dimensions, its properties may not be efficiently contractible. Contraction of a PEPS state is provably $\sharp$P hard\cite{schuch2007computational}. 
However, a physically relevant subset of these states can be efficiently contractible and an isometric representation of them (isometric PEPS for which the Moses move of Ref.\cite{zaletel2020isometric} can be carried out without approximation are indeed quasi-local and finitely contractible) could confer quantum advantage from the shallow representation of the consitiuent unitaries\cite{schwarz2012preparing}. The balance of advantage and cost in quantum algorithms can be delicate; extracting the elements of the tensors is easy classically, but quantum mechanically requires tomography of the circuit state, which is exponentially slow in the number of spins measured. This may be the bottleneck in hybrid algorithms\cite{wecker2015progress}. Using the tensor network framework to distribute entanglement resources over the Hilbert space appropriately can mitigate some of these costs.

This work demonstrates the utility of translating tensor network algorithms to quantum circuits and opens an unexplored direction for quantum simulation. Potentially all of the advances of classical simulation of quantum systems using tensor networks can be translated in this way. Moreover, it provides a complementary perspective on classical algorithms suggesting related benefit in purely unitary implementations\cite{Green_Pollmann_Unpublished}. 

\section{Methods}
\label{app:Methods}

\subsection{Quantum Matrix Product States}
\label{app:Appendix1}
The mapping from MPS to quantum circuits that we use automatically embodies much of the variational manifold and its tangent space. The parsimony of this mapping to the quantum circuit suggests that it is the natural home for MPS. The fundamental building block of the circuits depicted in Figs. \ref{fig:QiMPS}, \ref{fig:QTDVP} is the MPS tensor. 
A tensor of bond order $D$ and local Hilbert space dimension $D$ is represented by an $SU(dD)$ matrix following \cite{schon2005sequential,mps_path_integrals, huggins2019towards, sarang_finite_depth, ran2020encoding} $A^\sigma_{ij} = U_{(1 \otimes j),(\sigma \otimes i)}$ as shown in Fig.{\ref{appfig:MPSvsU}. This translation automatically encodes the left canonical form of the MPS tensor; $\sum_\sigma (A^\sigma)_{ij}^\dagger A^\sigma_{jk}= \delta_{ik}$. This follows directly from the unitary property of $U$. A classical implementation of an MPS algorithm involves returning the tensors to this form after each step in an algorithm using singular value decompositions --- in a quantum algorithm, such a manipulation is not required. 

Moreover, the remaining elements of the unitary encode the tangent space structure to the sub-manifold of states spanded by MPS. These are important in constructing the projected Hamiltonian dynamics. Adopting the notation of Ref.\cite{haegemanTDVP}, $V_{(\sigma \otimes \delta\ne 1),(i \otimes j)} = U_{(\delta\ne 1 \otimes j),(\sigma\otimes i)}$
and automatically satisfies the null or tangent gauge-fixing condition $\sum_\sigma (A^\sigma)_{ij}^\dagger V^{\sigma, \delta \ne1}_{jk}= 0$. This structure is responsible for the very compact quantum  implementation of the time-dependent variational principle shown in Fig. \ref{fig:QTDVP}. It obviates the need to calculate the tangent space structure at each step\cite{haegemanTDVP}.

 \begin{figure}
     \includegraphics[width=\linewidth]{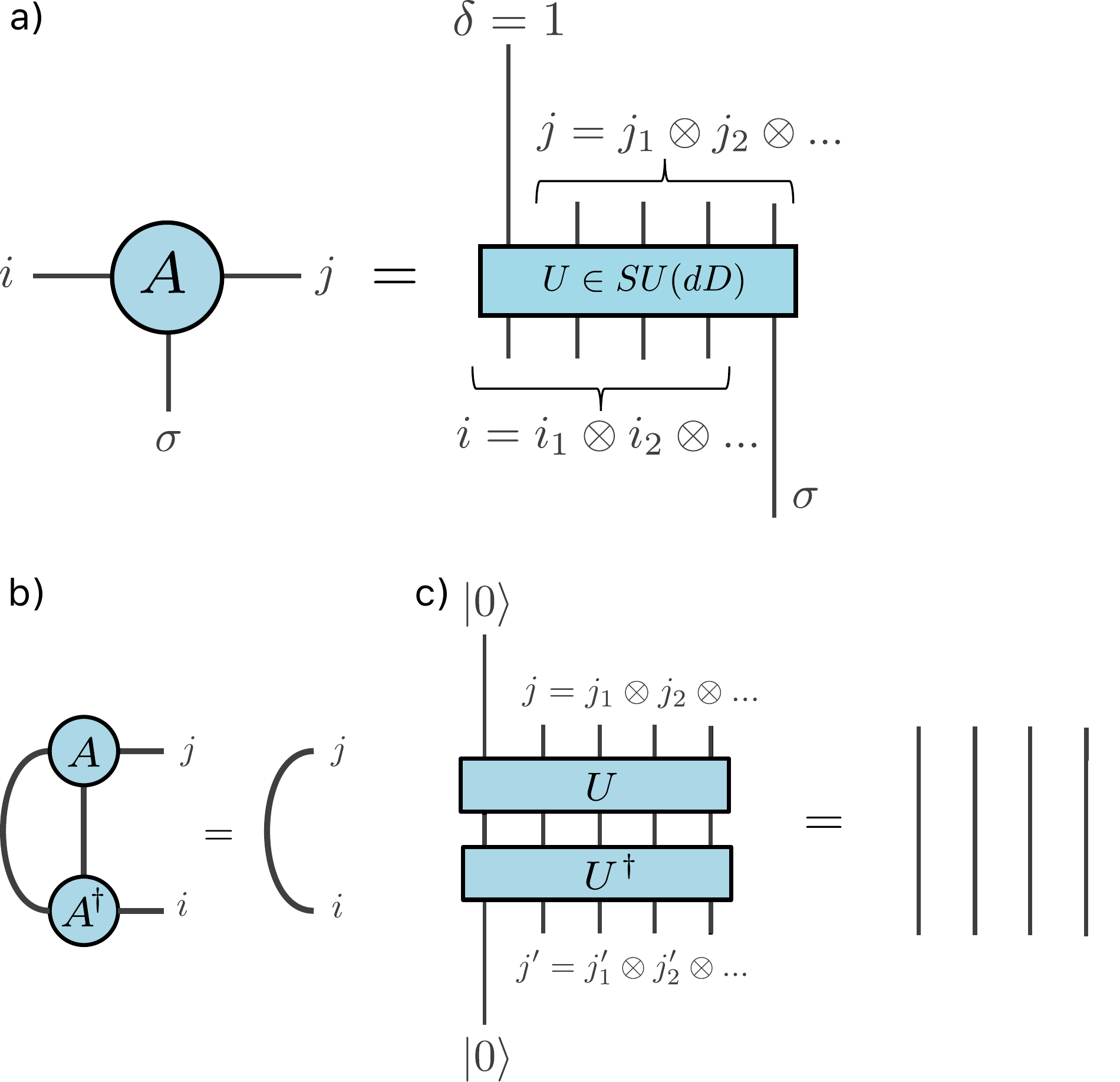}
\caption{{\bf Translation between MPS quantum circuits:}  a) The translation of an MPS tensor to a quantum circuit. The auxilliary index of bond order $D$ is created from $N= \log_2 D$ qubits.  b) The canonical form implies that the effect of contracting the MPS to the left is trivial. c) With our mapping of the MPS to a quantum circuit, the unitarity of $U$ automatically puts the tensor in left canonical form.
}
\label{appfig:MPSvsU}
\end{figure}

\subsection{The Quantum Time-dependent Variational Principle}
\label{app:Appendix2}
The equivalence with the classical implementation of the time-dependent variational principle for matrix product states and its quantum version can be seen by adopting the following parametrization of the updated unitary in the form 
$$
U^\prime = 
U \exp \left( \begin{array}{cc} 0 & X^\dagger \\ X & 0 \end{array} \right) .
$$
Taking the explicit overlap of the circuit in Fig. \ref{fig:QTDVP} a) with the state $|000 ... \rangle$ and then calculating its derivative with respect to $X$ recovers the recovers the time-dependent variational principle as formulated in Ref.~\cite{haegemanTDVP}. The tensor $X$ is to be compared with that in Ref.~\cite{haegemanTDVP} rescaled by the square root of the environment tensor. The quickest route to see this is to expand the circuit to quadratic order in the tensor $X$ and bi-linear order in $X$ and $dt$, before differentiating with respect to $X$. 

\subsection{Optimising Quantum Circuits}
\label{app:Appendix3}
Our algorithms require the optimisation of expectations of observables --- in Figs. \ref{fig:QiMPS}b), and \ref{fig:QTDVP}a) --- and the solution of fixed point equations in Figs. \ref{fig:QiMPS}b), and \ref{fig:QTDVP}b) and c) to determine the environment and mixed environment. In all cases, optimisations are carried out stochastically. 

We use the Rotosolve algorithm\cite{Ostaszewski2021structure} to speed up our stochastic searches. This utilises the fact that the dependence of expectations of a parametrized quantum circuit on any particular parameter is sinusoidal. As a result, after just three measurements one can take this parameter to its local optimum value. Extensions of this allow the variation to be calculated when several elements of the circuit depend upon same parameter. 

The equations illustrated in Figs. \ref{fig:QiMPS}b), and \ref{fig:QTDVP}b) and c) are implicitly identities between density matrices. We solve them using a version of the swap test that amounts to a stochastic optimisation of the objective function $tr \left[(\hat r - \hat s)^\dagger (\hat r - \hat s) \right]$.Details of how the swap test is implemented for the environment in Fig. \ref{fig:QiMPS}b) and for the mixed environments in Figs. \ref{fig:QTDVP}b) and c) are given in the supplementary materials. 

\section{Data Availability Statement}
All data generated or analysed during this study are included in this published article (and its supplementary information files).

\section{Code Availability Statement}
All code used to generate the results presented is available to download at https://github.com/fergusbarratt/qmps. We provide both Cirq code --- which we have used in our simulations --- and Qiskit code with the same functionality. 

\section{Acknowledgements}

JD, FB and AGG were supported by the EPSRC through grants EP/L015242/1, EP/L015854/1 and EP/S005021/1.  
FP is funded by the European Research Council (ERC) under the European Union's Horizon 2020 research and innovation program (grant agreement No. 771537). FP acknowledges the support of the DFG Research Unit FOR 1807 through grants no.PO 1370/2-1, TRR80, and the Deutsche Forschungsgemeinschaft (DFG, German Research Foundation) under Germany's Excellence Strategy EXC-2111-390814868.
We would like to acknowledge discussions with Adam Smith and Bernard Jobst, and GoogleAI for supporting attendance of FB at a Cirq coding workshop.

\section{Author Contributions}
AGG conceived the project and developed it initially with VS and MB.  FB, JD and AGG worked out the detailed mapping of these ideas to a quantum circuit, wrote the Cirq and Qiskit code and ran the simulations. FP made key contributions at various points in the code development. The manuscript was written by FB, JD and AGG.

\section{Competing Interests}
The Authors declare no competing interests. 

\bibliographystyle{naturemag}
\bibliography{BibliographySubmission.bib}

\begin{thebibliography}{10}
\expandafter\ifx\csname url\endcsname\relax
  \def\url#1{\texttt{#1}}\fi
\expandafter\ifx\csname urlprefix\endcsname\relax\def\urlprefix{URL }\fi
\providecommand{\bibinfo}[2]{#2}
\providecommand{\eprint}[2][]{\url{#2}}

\bibitem{SteveWhiteDMRG}
\bibinfo{author}{White, S.~R.}
\newblock \bibinfo{title}{Density matrix formulation for quantum
  renormalization groups}.
\newblock \emph{\bibinfo{journal}{Phys. Rev. Lett.}}
  \textbf{\bibinfo{volume}{69}}, \bibinfo{pages}{2863--2866}
  (\bibinfo{year}{1992}).

\bibitem{OstlundRommer}
\bibinfo{author}{{\"O}stlund, S.} \& \bibinfo{author}{Rommer, S.}
\newblock \bibinfo{title}{Thermodynamic limit of density matrix
  renormalization}.
\newblock \emph{\bibinfo{journal}{Phys. Rev. Lett.}}
  \textbf{\bibinfo{volume}{75}}, \bibinfo{pages}{3537--3540}
  (\bibinfo{year}{1995}).

\bibitem{mps_representations}
\bibinfo{author}{Perez-Garcia, D.}, \bibinfo{author}{Verstraete, F.},
  \bibinfo{author}{Wolf, M.~M.} \& \bibinfo{author}{Cirac, J.~I.}
\newblock \bibinfo{title}{Matrix product state representations}.
\newblock \emph{\bibinfo{journal}{Quantum Info. Comput.}}
  \textbf{\bibinfo{volume}{7}}, \bibinfo{pages}{401--430}
  (\bibinfo{year}{2007}).

\bibitem{mps_classically_efficient}
\bibinfo{author}{Vidal, G.}
\newblock \bibinfo{title}{Efficient classical simulation of slightly entangled
  quantum computations}.
\newblock \emph{\bibinfo{journal}{Phys. Rev. Lett.}}
  \textbf{\bibinfo{volume}{91}}, \bibinfo{pages}{147902}
  (\bibinfo{year}{2003}).

\bibitem{VQE}
\bibinfo{author}{Peruzzo, A.} \emph{et~al.}
\newblock \bibinfo{title}{A variational eigenvalue solver on a photonic quantum
  processor}.
\newblock \emph{\bibinfo{journal}{Nat. Commun.}} \textbf{\bibinfo{volume}{5}},
  \bibinfo{pages}{4213} (\bibinfo{year}{2014}).
\newblock \urlprefix\url{https://doi.org/10.1038/ncomms5213}.

\bibitem{phase_estimation}
\bibinfo{author}{Kitaev, A.~Y.}
\newblock \bibinfo{title}{Quantum measurements and the abelian stabilizer
  problem}  (\bibinfo{year}{1995}).
\newblock \eprint{arXiv:quant-ph/9511026}.

\bibitem{trotter1959product}
\bibinfo{author}{Trotter, H.~F.}
\newblock \bibinfo{title}{On the product of semi-groups of operators}.
\newblock \emph{\bibinfo{journal}{Proc. Am. Math. Soc.}}
  \textbf{\bibinfo{volume}{10}}, \bibinfo{pages}{545--551}
  (\bibinfo{year}{1959}).

\bibitem{SUZUKI1993232}
\bibinfo{author}{Suzuki, M.}
\newblock \bibinfo{title}{Improved trotter-like formula}.
\newblock \emph{\bibinfo{journal}{Phys. Lett. A}}
  \textbf{\bibinfo{volume}{180}}, \bibinfo{pages}{232 -- 234}
  (\bibinfo{year}{1993}).
\newblock
  \urlprefix\url{http://www.sciencedirect.com/science/article/pii/037596019390701Z}.

\bibitem{hastings2014improving}
\bibinfo{author}{Hastings, M.~B.}, \bibinfo{author}{Wecker, D.},
  \bibinfo{author}{Bauer, B.} \& \bibinfo{author}{Troyer, M.}
\newblock \bibinfo{title}{Improving quantum algorithms for quantum chemistry}.
\newblock \emph{\bibinfo{journal}{arXiv preprint arXiv:1403.1539}}
  (\bibinfo{year}{2014}).

\bibitem{BenjaminTDVP}
\bibinfo{author}{Li, Y.} \& \bibinfo{author}{Benjamin, S.~C.}
\newblock \bibinfo{title}{Efficient variational quantum simulator incorporating
  active error minimization}.
\newblock \emph{\bibinfo{journal}{Phys. Rev. X}} \textbf{\bibinfo{volume}{7}},
  \bibinfo{pages}{021050} (\bibinfo{year}{2017}).

\bibitem{circuits_are_mps}
\bibinfo{author}{Sch\"on, C.}, \bibinfo{author}{Hammerer, K.},
  \bibinfo{author}{Wolf, M.~M.}, \bibinfo{author}{Cirac, J.~I.} \&
  \bibinfo{author}{Solano, E.}
\newblock \bibinfo{title}{Sequential generation of matrix-product states in
  cavity qed}.
\newblock \emph{\bibinfo{journal}{Phys. Rev. A}} \textbf{\bibinfo{volume}{75}},
  \bibinfo{pages}{032311} (\bibinfo{year}{2007}).
\newblock \urlprefix\url{https://link.aps.org/doi/10.1103/PhysRevA.75.032311}.

\bibitem{haegemanTDVP}
\bibinfo{author}{Haegeman, J.} \emph{et~al.}
\newblock \bibinfo{title}{Time-dependent variational principle for quantum
  lattices}.
\newblock \emph{\bibinfo{journal}{Phys. Rev. Lett.}}
  \textbf{\bibinfo{volume}{107}}, \bibinfo{pages}{070601}
  (\bibinfo{year}{2011}).
\newblock
  \urlprefix\url{https://link.aps.org/doi/10.1103/PhysRevLett.107.070601}.

\bibitem{mps_path_integrals}
\bibinfo{author}{Green, A.~G.}, \bibinfo{author}{Hooley, C.~A.},
  \bibinfo{author}{Keeling, J.} \& \bibinfo{author}{Simon, S.~H.}
\newblock \bibinfo{title}{Feynman path integrals over entangled states}
  (\bibinfo{year}{2016}).
\newblock \eprint{Preprint at http://arXiv.org/abs/1607.01778}.

\bibitem{peng2020simulating}
\bibinfo{author}{Peng, T.}, \bibinfo{author}{Harrow, A.~W.},
  \bibinfo{author}{Ozols, M.} \& \bibinfo{author}{Wu, X.}
\newblock \bibinfo{title}{Simulating large quantum circuits on a small quantum
  computer}.
\newblock \emph{\bibinfo{journal}{Phys. Rev. Lett.}}
  \textbf{\bibinfo{volume}{125}}, \bibinfo{pages}{150504}
  (\bibinfo{year}{2020}).

\bibitem{kim2017holographic}
\bibinfo{author}{Kim, I.~H.}
\newblock \bibinfo{title}{Holographic quantum simulation}.
\newblock \emph{\bibinfo{journal}{Preprint at
  http://arXiv.org/quantu-ph/1702.02093}}  (\bibinfo{year}{2017}).

\bibitem{kim2017noise}
\bibinfo{author}{Kim, I.~H.}
\newblock \bibinfo{title}{Noise-resilient preparation of quantum many-body
  ground states}.
\newblock \emph{\bibinfo{journal}{Preprint at
  https://arxiv.org/abs/1703.00032}}  (\bibinfo{year}{2017}).

\bibitem{lin2021real}
\bibinfo{author}{Lin, S.-H.}, \bibinfo{author}{Dilip, R.},
  \bibinfo{author}{Green, A.~G.}, \bibinfo{author}{Smith, A.} \&
  \bibinfo{author}{Pollmann, F.}
\newblock \bibinfo{title}{Real- and imaginary-time evolution with compressed
  quantum circuits}.
\newblock \emph{\bibinfo{journal}{PRX Quantum}} \textbf{\bibinfo{volume}{2}},
  \bibinfo{pages}{010342} (\bibinfo{year}{2021}).
\newblock \urlprefix\url{https://link.aps.org/doi/10.1103/PRXQuantum.2.010342}.

\bibitem{schollwock_review}
\bibinfo{author}{Schollw{\"{o}}ck, U.}
\newblock \bibinfo{title}{The density-matrix renormalization group in the age
  of matrix product states}.
\newblock \emph{\bibinfo{journal}{Ann. Phys. (NY)}}
  \textbf{\bibinfo{volume}{326}}, \bibinfo{pages}{96 -- 192}
  (\bibinfo{year}{2011}).
\newblock
  \urlprefix\url{http://www.sciencedirect.com/science/article/pii/S0003491610001752}.
\newblock \bibinfo{note}{January 2011 Special Issue}.

\bibitem{orus_review}
\bibinfo{author}{Or{\'{u}}s, R.}
\newblock \bibinfo{title}{A practical introduction to tensor networks: Matrix
  product states and projected entangled pair states}.
\newblock \emph{\bibinfo{journal}{Ann. Phys. (NY)}}
  \textbf{\bibinfo{volume}{349}}, \bibinfo{pages}{117 -- 158}
  (\bibinfo{year}{2014}).
\newblock
  \urlprefix\url{http://www.sciencedirect.com/science/article/pii/S0003491614001596}.

\bibitem{Smith2019}
\bibinfo{author}{Smith, A.}, \bibinfo{author}{Jobst, B.},
  \bibinfo{author}{Green, A.~G.} \& \bibinfo{author}{Pollmann, F.}
\newblock \bibinfo{title}{Crossing a topological phase transition with a
  quantum computer} \eprint{Preprint at http://arxiv.org/abs/1910.05351v2}.

\bibitem{nielsen_chuang}
\bibinfo{author}{Nielsen, M.~A.} \& \bibinfo{author}{Chuang, I.~L.}
\newblock \emph{\bibinfo{title}{Quantum Computation and Quantum Information:
  10th Anniversary Edition}} (\bibinfo{publisher}{Cambridge University Press},
  \bibinfo{address}{USA}, \bibinfo{year}{2011}), \bibinfo{edition}{10th} edn.

\bibitem{cerezo2021cost}
\bibinfo{author}{Cerezo, M.}, \bibinfo{author}{Sone, A.},
  \bibinfo{author}{Volkoff, T.}, \bibinfo{author}{Cincio, L.} \&
  \bibinfo{author}{Coles, P.~J.}
\newblock \bibinfo{title}{Cost function dependent barren plateaus in shallow
  parametrized quantum circuits}.
\newblock \emph{\bibinfo{journal}{Nat. Commun.}} \textbf{\bibinfo{volume}{12}},
  \bibinfo{pages}{1--12} (\bibinfo{year}{2021}).

\bibitem{vatan2004optimal}
\bibinfo{author}{Vatan, F.} \& \bibinfo{author}{Williams, C.}
\newblock \bibinfo{title}{Optimal quantum circuits for general two-qubit
  gates}.
\newblock \emph{\bibinfo{journal}{Phys. Rev. A}} \textbf{\bibinfo{volume}{69}},
  \bibinfo{pages}{032315} (\bibinfo{year}{2004}).

\bibitem{dynamical_phase_transitions}
\bibinfo{author}{Heyl, M.}, \bibinfo{author}{Polkovnikov, A.} \&
  \bibinfo{author}{Kehrein, S.}
\newblock \bibinfo{title}{Dynamical quantum phase transitions in the
  transverse-field ising model}.
\newblock \emph{\bibinfo{journal}{Phys. Rev. Lett.}}
  \textbf{\bibinfo{volume}{110}}, \bibinfo{pages}{135704}
  (\bibinfo{year}{2013}).
\newblock
  \urlprefix\url{https://link.aps.org/doi/10.1103/PhysRevLett.110.135704}.

\bibitem{michailidis2020slow}
\bibinfo{author}{Michailidis, A.}, \bibinfo{author}{Turner, C.},
  \bibinfo{author}{Papi{\'c}, Z.}, \bibinfo{author}{Abanin, D.} \&
  \bibinfo{author}{Serbyn, M.}
\newblock \bibinfo{title}{Slow quantum thermalization and many-body revivals
  from mixed phase space}.
\newblock \emph{\bibinfo{journal}{Phys. Rev. X}} \textbf{\bibinfo{volume}{10}},
  \bibinfo{pages}{011055} (\bibinfo{year}{2020}).

\bibitem{turner2018weak}
\bibinfo{author}{Turner, C.~J.}, \bibinfo{author}{Michailidis, A.~A.},
  \bibinfo{author}{Abanin, D.~A.}, \bibinfo{author}{Serbyn, M.} \&
  \bibinfo{author}{Papi{\'c}, Z.}
\newblock \bibinfo{title}{Weak ergodicity breaking from quantum many-body
  scars}.
\newblock \emph{\bibinfo{journal}{Nat. Phys.}} \textbf{\bibinfo{volume}{14}},
  \bibinfo{pages}{745--749} (\bibinfo{year}{2018}).

\bibitem{zaletel2020isometric}
\bibinfo{author}{Zaletel, M.~P.} \& \bibinfo{author}{Pollmann, F.}
\newblock \bibinfo{title}{Isometric tensor network states in two dimensions}.
\newblock \emph{\bibinfo{journal}{Phys. Rev. Lett.}}
  \textbf{\bibinfo{volume}{124}}, \bibinfo{pages}{037201}
  (\bibinfo{year}{2020}).

\bibitem{mera_vidal}
\bibinfo{author}{Vidal, G.}
\newblock \bibinfo{title}{Class of quantum many-body states that can be
  efficiently simulated}.
\newblock \emph{\bibinfo{journal}{Phys. Rev. Lett.}}
  \textbf{\bibinfo{volume}{101}}, \bibinfo{pages}{110501}
  (\bibinfo{year}{2008}).
\newblock
  \urlprefix\url{https://link.aps.org/doi/10.1103/PhysRevLett.101.110501}.

\bibitem{kim2017robust}
\bibinfo{author}{Kim, I.~H.} \& \bibinfo{author}{Swingle, B.}
\newblock \bibinfo{title}{Robust entanglement renormalization on a noisy
  quantum computer}.
\newblock \emph{\bibinfo{journal}{Preprint at
  https://arxiv.org/abs/1711.07500}}  (\bibinfo{year}{2017}).

\bibitem{hierarchical_quantum_classifiers}
\bibinfo{author}{Grant, E.} \emph{et~al.}
\newblock \bibinfo{title}{Hierarchical quantum classifiers}.
\newblock \emph{\bibinfo{journal}{npj Quantum Inf.}}
  \textbf{\bibinfo{volume}{4}}, \bibinfo{pages}{65} (\bibinfo{year}{2018}).
\newblock \urlprefix\url{https://doi.org/10.1038/s41534-018-0116-9}.

\bibitem{cong2019quantum}
\bibinfo{author}{Cong, I.}, \bibinfo{author}{Choi, S.} \&
  \bibinfo{author}{Lukin, M.~D.}
\newblock \bibinfo{title}{Quantum convolutional neural networks}.
\newblock \emph{\bibinfo{journal}{Nat. Phys.}} \textbf{\bibinfo{volume}{15}},
  \bibinfo{pages}{1273--1278} (\bibinfo{year}{2019}).

\bibitem{schuch2007computational}
\bibinfo{author}{Schuch, N.}, \bibinfo{author}{Wolf, M.~M.},
  \bibinfo{author}{Verstraete, F.} \& \bibinfo{author}{Cirac, J.~I.}
\newblock \bibinfo{title}{Computational complexity of projected entangled pair
  states}.
\newblock \emph{\bibinfo{journal}{Phys. Rev. Lett.}}
  \textbf{\bibinfo{volume}{98}}, \bibinfo{pages}{140506}
  (\bibinfo{year}{2007}).

\bibitem{schwarz2012preparing}
\bibinfo{author}{Schwarz, M.}, \bibinfo{author}{Temme, K.} \&
  \bibinfo{author}{Verstraete, F.}
\newblock \bibinfo{title}{Preparing projected entangled pair states on a
  quantum computer}.
\newblock \emph{\bibinfo{journal}{Phys. Rev. Lett.}}
  \textbf{\bibinfo{volume}{108}}, \bibinfo{pages}{110502}
  (\bibinfo{year}{2012}).

\bibitem{wecker2015progress}
\bibinfo{author}{Wecker, D.}, \bibinfo{author}{Hastings, M.~B.} \&
  \bibinfo{author}{Troyer, M.}
\newblock \bibinfo{title}{Progress towards practical quantum variational
  algorithms}.
\newblock \emph{\bibinfo{journal}{Phys. Rev. A}} \textbf{\bibinfo{volume}{92}},
  \bibinfo{pages}{042303} (\bibinfo{year}{2015}).

\bibitem{Green_Pollmann_Unpublished}
\bibinfo{author}{Green, A.~G.} \& \bibinfo{author}{Pollmann, F.}
\newblock \bibinfo{title}{Unpublished result.}  (\bibinfo{year}{2020}).

\bibitem{schon2005sequential}
\bibinfo{author}{Sch{\"o}n, C.}, \bibinfo{author}{Solano, E.},
  \bibinfo{author}{Verstraete, F.}, \bibinfo{author}{Cirac, J.~I.} \&
  \bibinfo{author}{Wolf, M.~M.}
\newblock \bibinfo{title}{Sequential generation of entangled multiqubit
  states}.
\newblock \emph{\bibinfo{journal}{Phys. Rev. Lett.}}
  \textbf{\bibinfo{volume}{95}}, \bibinfo{pages}{110503}
  (\bibinfo{year}{2005}).

\bibitem{huggins2019towards}
\bibinfo{author}{Huggins, W.}, \bibinfo{author}{Patil, P.},
  \bibinfo{author}{Mitchell, B.}, \bibinfo{author}{Whaley, K.~B.} \&
  \bibinfo{author}{Stoudenmire, E.~M.}
\newblock \bibinfo{title}{Towards quantum machine learning with tensor
  networks}.
\newblock \emph{\bibinfo{journal}{Quantum Sci. Technol.}}
  \textbf{\bibinfo{volume}{4}}, \bibinfo{pages}{024001} (\bibinfo{year}{2019}).
\newblock \urlprefix\url{https://doi.org/10.1088%2F2058-9565%2Faaea94}.

\bibitem{sarang_finite_depth}
\bibinfo{author}{Gopalakrishnan, S.} \& \bibinfo{author}{Lamacraft, A.}
\newblock \bibinfo{title}{Unitary circuits of finite depth and infinite width
  from quantum channels}.
\newblock \emph{\bibinfo{journal}{Phys. Rev. B}}
  \textbf{\bibinfo{volume}{100}}, \bibinfo{pages}{064309}
  (\bibinfo{year}{2019}).
\newblock \urlprefix\url{https://link.aps.org/doi/10.1103/PhysRevB.100.064309}.

\bibitem{ran2020encoding}
\bibinfo{author}{Ran, S.-J.}
\newblock \bibinfo{title}{Encoding of matrix product states into quantum
  circuits of one-and two-qubit gates}.
\newblock \emph{\bibinfo{journal}{Phys. Rev. A}}
  \textbf{\bibinfo{volume}{101}}, \bibinfo{pages}{032310}
  (\bibinfo{year}{2020}).

\bibitem{Ostaszewski2021structure}
\bibinfo{author}{Ostaszewski, M.}, \bibinfo{author}{Grant, E.} \&
  \bibinfo{author}{Benedetti, M.}
\newblock \bibinfo{title}{Structure optimization for parameterized quantum
  circuits}.
\newblock \emph{\bibinfo{journal}{{Quantum}}} \textbf{\bibinfo{volume}{5}},
  \bibinfo{pages}{391} (\bibinfo{year}{2021}).

\end{thebibliography}


\begin{thebibliography}{1}
\expandafter\ifx\csname url\endcsname\relax
  \def\url#1{\texttt{#1}}\fi
\expandafter\ifx\csname urlprefix\endcsname\relax\def\urlprefix{URL }\fi
\providecommand{\bibinfo}[2]{#2}
\providecommand{\eprint}[2][]{\url{#2}}

\bibitem{mps_representations}
\bibinfo{author}{Perez-Garcia, D.}, \bibinfo{author}{Verstraete, F.},
  \bibinfo{author}{Wolf, M.~M.} \& \bibinfo{author}{Cirac, J.~I.}
\newblock \bibinfo{title}{Matrix product state representations}.
\newblock \emph{\bibinfo{journal}{Quantum Info. Comput.}}
  \textbf{\bibinfo{volume}{7}}, \bibinfo{pages}{401--430}
  (\bibinfo{year}{2007}).

\bibitem{schollwock_review}
\bibinfo{author}{Schollw{\"{o}}ck, U.}
\newblock \bibinfo{title}{The density-matrix renormalization group in the age
  of matrix product states}.
\newblock \emph{\bibinfo{journal}{Ann. Phys. (NY)}}
  \textbf{\bibinfo{volume}{326}}, \bibinfo{pages}{96 -- 192}
  (\bibinfo{year}{2011}).
\newblock
  \urlprefix\url{http://www.sciencedirect.com/science/article/pii/S0003491610001752}.
\newblock \bibinfo{note}{January 2011 Special Issue}.

\bibitem{orus_review}
\bibinfo{author}{Or{\'{u}}s, R.}
\newblock \bibinfo{title}{A practical introduction to tensor networks: Matrix
  product states and projected entangled pair states}.
\newblock \emph{\bibinfo{journal}{Ann. Phys. (NY)}}
  \textbf{\bibinfo{volume}{349}}, \bibinfo{pages}{117 -- 158}
  (\bibinfo{year}{2014}).
\newblock
  \urlprefix\url{http://www.sciencedirect.com/science/article/pii/S0003491614001596}.

\bibitem{swap_test}
\bibinfo{author}{Garcia-Escartin, J.~C.} \& \bibinfo{author}{Chamorro-Posada,
  P.}
\newblock \bibinfo{title}{swap test and hong-ou-mandel effect are equivalent}.
\newblock \emph{\bibinfo{journal}{Phys. Rev. A}} \textbf{\bibinfo{volume}{87}},
  \bibinfo{pages}{052330} (\bibinfo{year}{2013}).
\newblock \urlprefix\url{https://link.aps.org/doi/10.1103/PhysRevA.87.052330}.

\bibitem{pollmann_symmetry}
\bibinfo{author}{Pollmann, F.} \& \bibinfo{author}{Turner, A.~M.}
\newblock \bibinfo{title}{Detection of symmetry-protected topological phases in
  one dimension}.
\newblock \emph{\bibinfo{journal}{Phys. Rev. B}} \textbf{\bibinfo{volume}{86}},
  \bibinfo{pages}{125441} (\bibinfo{year}{2012}).
\newblock \urlprefix\url{https://link.aps.org/doi/10.1103/PhysRevB.86.125441}.

\bibitem{Ostaszewski2021structure}
\bibinfo{author}{Ostaszewski, M.}, \bibinfo{author}{Grant, E.} \&
  \bibinfo{author}{Benedetti, M.}
\newblock \bibinfo{title}{Structure optimization for parameterized quantum
  circuits}.
\newblock \emph{\bibinfo{journal}{{Quantum}}} \textbf{\bibinfo{volume}{5}},
  \bibinfo{pages}{391} (\bibinfo{year}{2021}).

\bibitem{vidal2018calculus}
\bibinfo{author}{Vidal, J.~G.} \& \bibinfo{author}{Theis, D.~O.}
\newblock \bibinfo{title}{Calculus on parameterized quantum circuits}.
\newblock \emph{\bibinfo{journal}{arXiv preprint arXiv:1812.06323}}
  (\bibinfo{year}{2018}).

\bibitem{parrish2019jacobi}
\bibinfo{author}{Parrish, R.~M.}, \bibinfo{author}{Iosue, J.~T.},
  \bibinfo{author}{Ozaeta, A.} \& \bibinfo{author}{McMahon, P.~L.}
\newblock \bibinfo{title}{A jacobi diagonalization and anderson acceleration
  algorithm for variational quantum algorithm parameter optimization}.
\newblock \emph{\bibinfo{journal}{arXiv preprint arXiv:1904.03206}}
  (\bibinfo{year}{2019}).

\bibitem{nakanishi2020sequential}
\bibinfo{author}{Nakanishi, K.~M.}, \bibinfo{author}{Fujii, K.} \&
  \bibinfo{author}{Todo, S.}
\newblock \bibinfo{title}{Sequential minimal optimization for quantum-classical
  hybrid algorithms}.
\newblock \emph{\bibinfo{journal}{Physical Review Research}}
  \textbf{\bibinfo{volume}{2}}, \bibinfo{pages}{043158} (\bibinfo{year}{2020}).

\end{thebibliography}

\end{document}



\title{Supplementary Materials: Parallel Quantum Simulation of Large Systems on Small NISQ Computers}

\author{Fergus Barratt}
\affiliation{Department of Mathematics, King's College London, Strand, London WC2R 2LS, UK}
\author{James Dborin}
\affiliation{London Centre for Nanotechnology, University College London, Gordon St., London, WC1H 0AH, UK}
\author{Matthias Bal}
\affiliation{GTN Limited, Clifton House, 46 Clifton Terrace, Finsbury Park, London N4 3JP, UK}
\author{Vid Stojevic}
\affiliation{GTN Limited, Clifton House, 46 Clifton Terrace, Finsbury Park, London N4 3JP, UK}
\author{Frank Pollmann}
\affiliation{Department of Physics, T42, Technische Universit\"at M\"unchen, James-Franck-Stra{\ss}e 1, D-85748 Garching, Germany}
\author{A.~G. Green}
\affiliation{London Centre for Nanotechnology, University College London, Gordon St., London, WC1H 0AH, UK}

\date{\today}
\begin{abstract}
In this supplementary material we further explain the methods used in the quantum circuit implementation of  iMPS algorithms, and the effects of imperfections in that realisation. 
We begin with a summary of the main ideas of MPS and how they translate to quantum circuits. After this, we show how these formal circuit equations can be realised in practice.
We give details of appropriate ans\"atze and an optimization method that we have found effective. Next we detail the natural generalisation to higher bond order matrix product states, before giving additional data on the effects of shallow circuit realisations and finite gate fidelity. 
\end{abstract}
\maketitle

\section{Supplementary Methods}

\subsection{Quantum Circuit Realisations of MPS}
Matrix product state (MPS) techniques are now canonical in the condensed matter physics community. There are many fine reviews\cite{mps_representations,schollwock_review,orus_review}. Here we give a condensed summary of the additional details relevant to this work and how they translate to quantum circuits.

\subsubsection{Matrix Product States}
An MPS state for a spin-$1/2$ chain is written
\begin{eqnarray}
| \psi \rangle
&=&
\sum_{ \{ \sigma_\alpha \} } \sum_{ \{ i_\alpha \} } 
A^{\sigma_1}_{i_1} A^{\sigma_2}_{i_1,i_2} A^{\sigma_3}_{i_2,i_3}  ... A^{\sigma_N}_{i_N} 
| \sigma _1, \sigma_2, \sigma_3, ... , \sigma_N\rangle
\nonumber\\
&=& 
\;\;
\includegraphics[scale=0.65,valign=c]{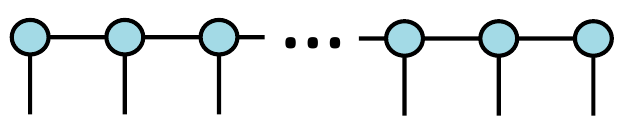}
\label{eq:MPSstate}
\end{eqnarray}
where we use a penrose diagrammatic representation in the second line\cite{mps_representations,schollwock_review,orus_review}. The tensor indices  $i_1, i_2, ... i_N$ run over values $1 \rightarrow D_i$. for a spin -$1/2$ system these states cover the Hilbert space when $D_i=2^{min(i, N-i)}$. In practice, $D_i$ is usually truncated at some value. 

\vspace{0.1in}
\noindent 
{\it Expectations of local operators} involve contracting the state along the whole chain and are nonlocal;
\begin{eqnarray}
\langle \psi | \hat \theta | \psi \rangle 
=
\includegraphics[scale=0.65,valign=c]{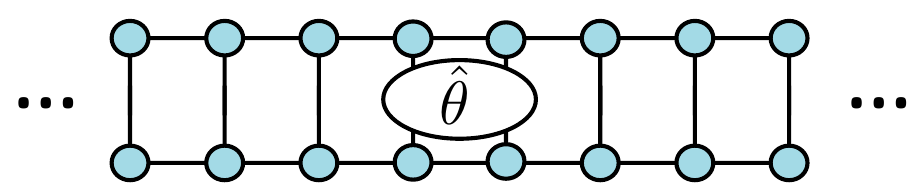}.
\label{eq:Expectation}
\end{eqnarray}
However, identifying the result of contracting everything to the left of the site $\alpha$ as $l_{\alpha}$ and to the right as $r_{\alpha}$ we may write this as a local expectation:
\begin{eqnarray}
& &
\langle \psi | \hat \theta | \psi \rangle 
=
\includegraphics[scale=0.6,valign=c]{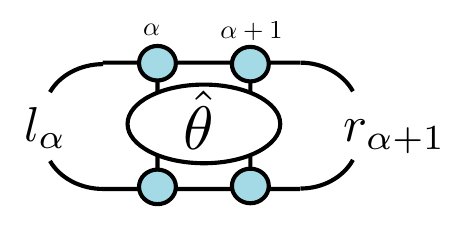},
\label{eq:LocalExpectation}
\\
& &
\includegraphics[scale=0.6,valign=c]{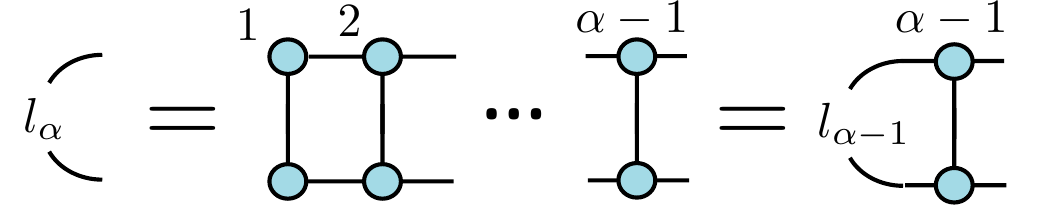},
\label{eq:LeftEnvironment}
\\
& &
\includegraphics[scale=0.6,valign=c]{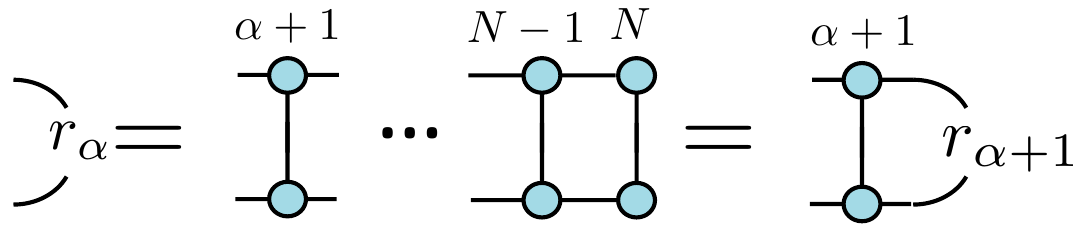},
\label{eq:RightEnvironment}
\end{eqnarray}
The tensors $l_\alpha$ and $r_\alpha$ are often called the environment for the site $\alpha$ and we will use that nomenclature here. 
Using a sequential singular-value decomposition from left to right of Eq.(\ref{eq:MPSstate}) allows the tensor to be put in a canonical form - or more precisely, left orthogonal form - such that $l_\alpha \equiv \bm{I}_{D_{i-1}\otimes D_{i-1}}$ $\forall \alpha$. This {\it isometric form} is the basis of our mapping to the quantum circuit shown in Fig.6. 

\subsubsection{Translation to Quantum Circuit}
A quantum circuit realisation of \SEq(\ref{eq:MPSstate}) is given by
\begin{eqnarray}
|\psi \rangle &=&
\includegraphics[scale=0.6,valign=c]{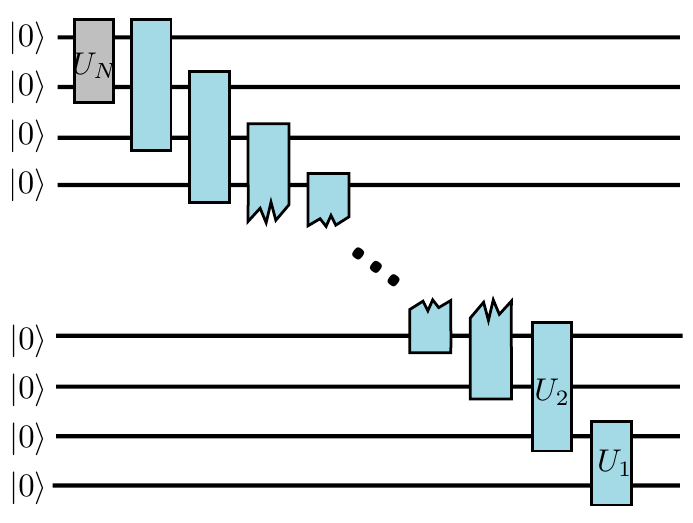}.
\label{eq:MPStoQMPS}
\end{eqnarray}
Notice by comparing with Fig.6a in the main paper that the bond order increases and reduces at the ends of the chain. We have truncated the bond order at $D=4$ in this equation. We note also that the greyed-out unitary at the end of the chain is redundant, {\it i.e.} it may be replaced with the identity. It represents the MPS tensors on the last site in this bond order $4$ case. There are $n-1$ such redundant unitaries at bond order $D=2^n$.

The circuits for the left environment $l_\alpha$ can be written
\begin{eqnarray}
l_{\alpha} 
&=&
\includegraphics[scale=0.5,valign=c]{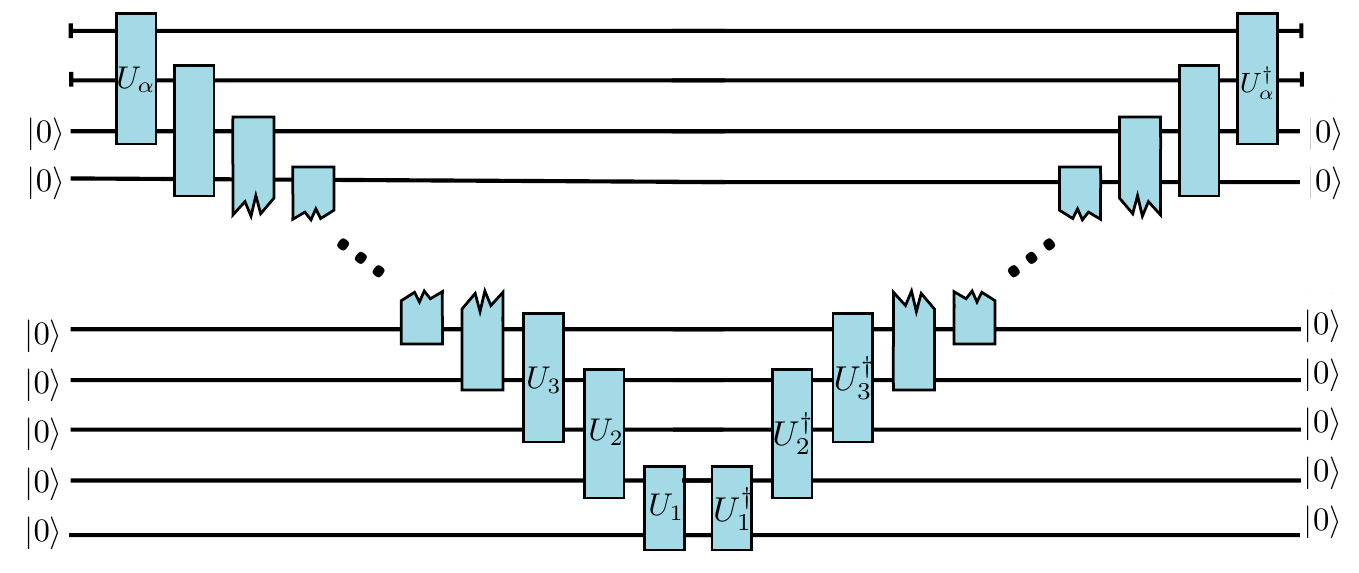}
\nonumber\\
&=&
\includegraphics[scale=0.5,valign=c]{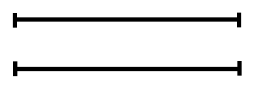}.
\label{eq:CircuitLeftEnvironment}
\end{eqnarray}
As indicated, the unitaries cancel in conjugate pairs and the left environment is trivial reflecting the left-orthogonal form of the MPS state that the circuit represents. 
The right environment can similarly be expressed by the following circuits
\begin{eqnarray}
r_\alpha
&=&
\includegraphics[scale=0.55,valign=c]{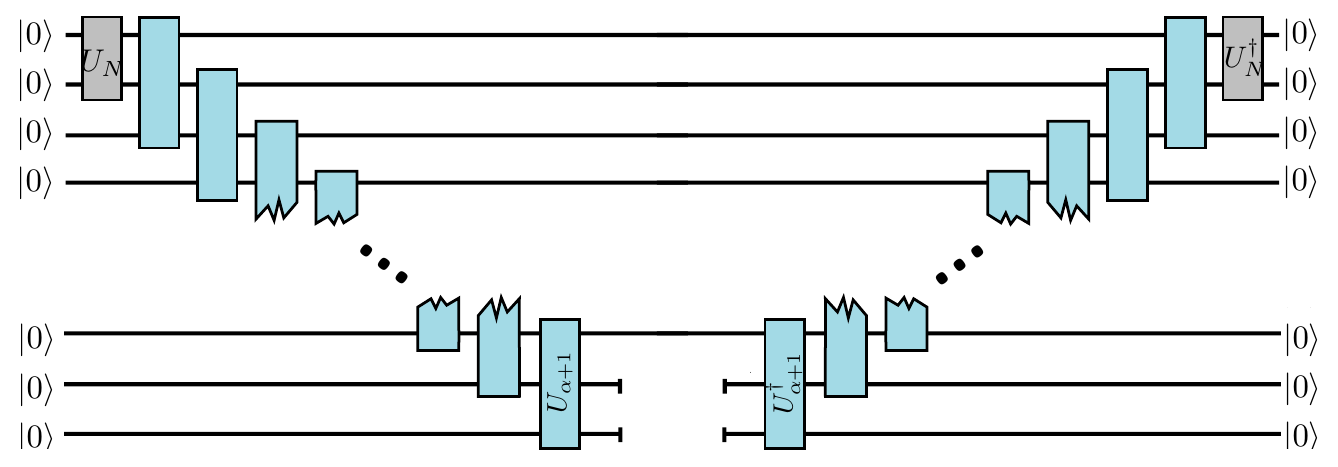}
\nonumber\\
&=&
\includegraphics[scale=0.6,valign=c]{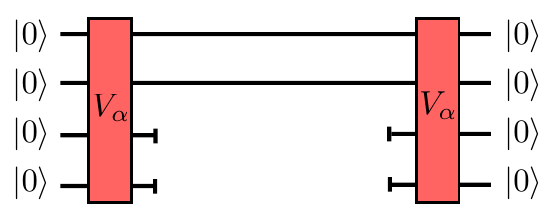}
=
\includegraphics[scale=0.6,valign=c]{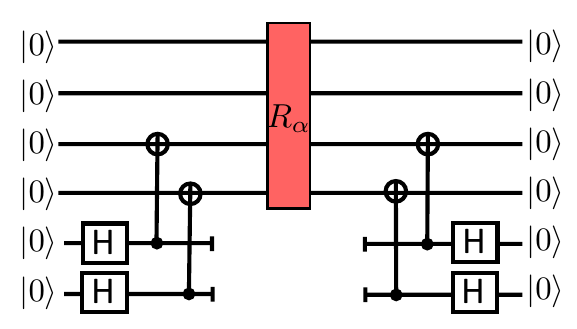}.
\nonumber\\
\label{eq:CircuitRightEnvironment}
\end{eqnarray}
The existence of the compressed tensors $V$ is the basis of our construction. It is guaranteed as follows: The right environment is Hermitian and has eigen-decomposition $r=U \Lambda U^\dagger$. 
Define $X = U\sqrt{\Lambda}$, and embed $X$ across the legs of the unitary $V$ by assigning $X$, reshaped into a vector, to the first column of $V$ - since $V$ is always applied to a state 
$|0\rangle^{\otimes N}$, these are the only entries of $V$ that feature in any calculation. The other columns of $V$ are chosen orthonormally and arbitrarily. 
We can choose to compress further by choosing only the $k$ highest eigenvalue eigenvector pairs to construct $U$ and $\Lambda$. 
The resulting environment $U'\Lambda'U^{\dagger'}$ is the rank-$k$ matrix that best approximates $r$ in Frobenius norm. This is the circuit realisation of the truncation scheme habitually applied in MPS calculations. In essence our work uses a variational scheme to determine $V$. 

A very similar argument can be used to produce the tensor $R$: The right environment $r$ is embedded in the appropriate diagonal block of $R$ the remaining elements - upon which no computation depends - are completed by an appropriate singular valued decomposition.

From \SEq(\ref{eq:CircuitRightEnvironment}) it is evident that we iteratively construct $r_\alpha$ or $V_\alpha$ from $r_{\alpha-1}$ or $V_{\alpha-1}$ according to
\begin{eqnarray}
\includegraphics[scale=0.6,valign=c]{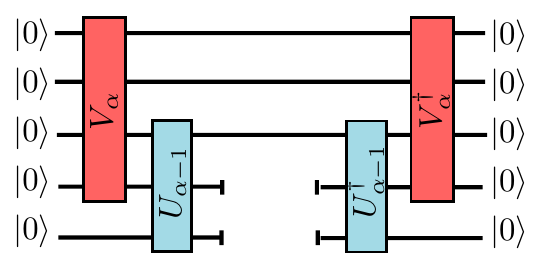}
=
\includegraphics[scale=0.6,valign=c]{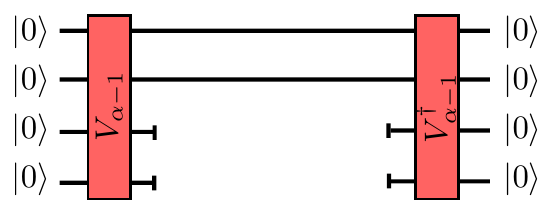}.
\label{eq:IterativeCircuitRightEnvironment}
\end{eqnarray}
or equivalently
\begin{eqnarray}
& &\includegraphics[scale=0.6,valign=c]{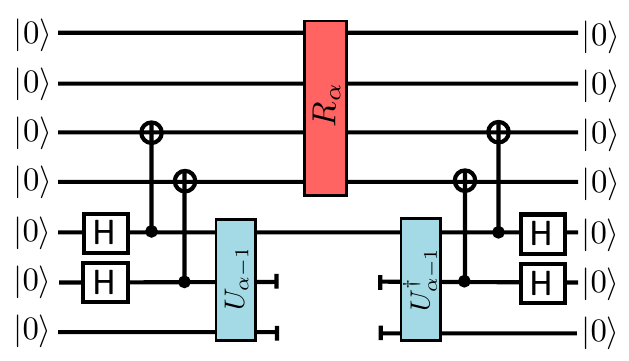}
=
\includegraphics[scale=0.6,valign=c]{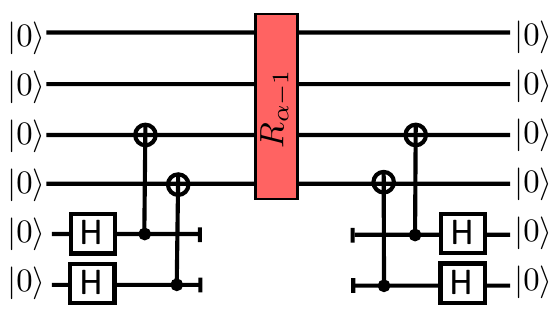}.
\nonumber\\
\label{eq:IterativeCircuitRightEnvironment}
\end{eqnarray}
These are the quantum circuit counterparts of the iterative expression for the MPS environment contained in the last part of \SEq(\ref{eq:RightEnvironment}). 
In {\it translationally invariant systems}, $V$ (and $R$) is independent of the site on the chain and \SEq(\ref{eq:IterativeCircuitRightEnvironment}) reduces to the fixed point equation seen in Fig.~2 c. 

\subsubsection{Time-evolving Quantum Circuit MPS}
The basis of our time-evolution algorithm is to find the state $|\psi (X(t+dt))\rangle$, parametrised by the set of parameters $X(t+dt)$, whose fidelity with the time-evolution of the state $|\psi (X(t))\rangle$ is optimised:t  
$$
\max_{X(t+dt)} \langle \psi (X(t+dt))| e^{i H dt} |\psi (X(t))\rangle|^2 .
$$
This overlap can be written for the $D=4$ quantum circuit MPS as 
\begin{eqnarray}
& &
 \langle \psi (X(t+dt))| e^{i H dt} |\psi (X(t))\rangle
 \nonumber\\
&=&\includegraphics[scale=0.6,valign=c]{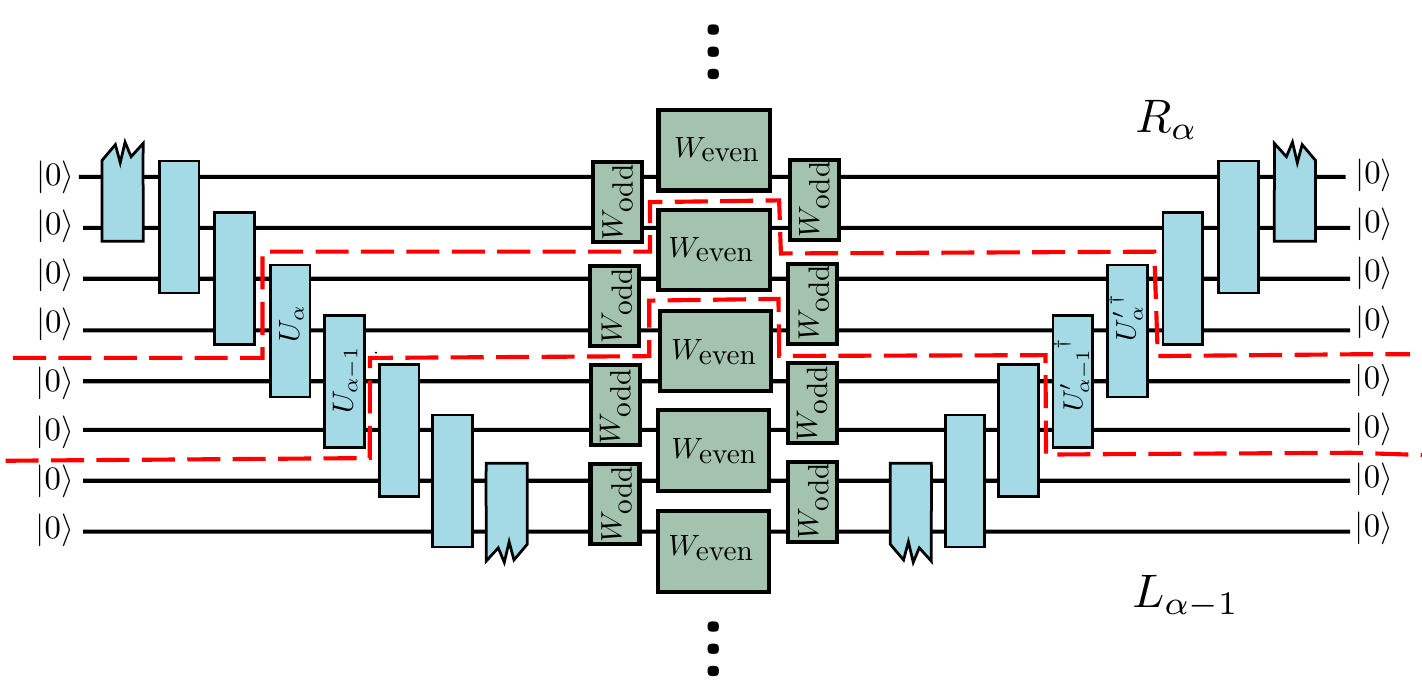},
\nonumber\\
\label{eq:EvolutionOverlap}
\end{eqnarray}
where we have expanded the time-evolution operator in the usual second order Trottterization as 
\begin{eqnarray}
\exp[{iHdt}]
&=&
\exp[{i H_{\hbox{odd}} dt/2}] 
\exp[  {i H_{\hbox{even}} dt}] 
\exp[{i H_{\hbox{odd}} dt/2}]
\nonumber\\
&=& 
W_{\hbox{odd}} W_{\hbox{even}} W_{\hbox{odd}} .
\end{eqnarray}
 The unitaries $U_\alpha$ encode the state at time $t$. When the modulus of this overlap is optimised, the unitaries $U'_\alpha$ encode the state at time $t+dt$. 

The circuit shown in \SEq(\ref{eq:EvolutionOverlap}) can be implemented on a quantum computer that is much smaller than the system under consideration by splitting along the red dashed lines. The resulting circuit takes the form
\begin{eqnarray}
& &
 \langle \psi (X(t+dt))| e^{i H dt} |\psi (X(t))\rangle
 \nonumber\\
&=&\includegraphics[scale=0.6,valign=c]{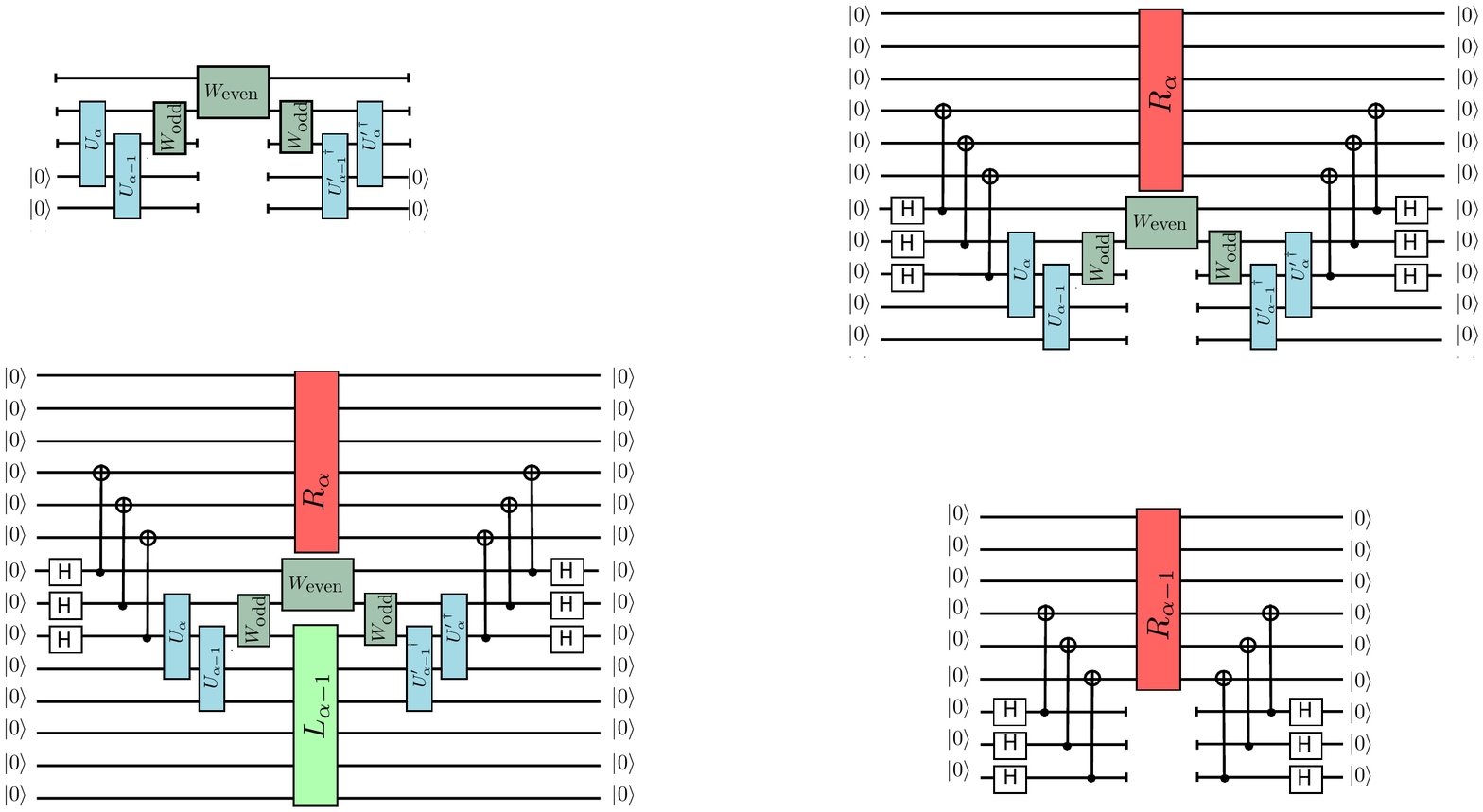},
\label{eq:EvolutionOverlapDivided}
\end{eqnarray}
where the mixed environment circuits (so called because $U$ and $U'$ are not the same) $R_\alpha$ and $L_\alpha$ are the portions of the circuit above and below the red dashed lines in \SEq(\ref{eq:EvolutionOverlap}), compressed to tensors $R_\alpha$ and $L\alpha$ as in \SEq(\ref{eq:CircuitRightEnvironment}). Following the same procedure as in \SEq(\ref{eq:IterativeCircuitRightEnvironment}), $R_\alpha$ and $L_\alpha$ can be determined iteratively from site to site from the equations
\begin{eqnarray}
& &
\includegraphics[scale=0.6,valign=c]{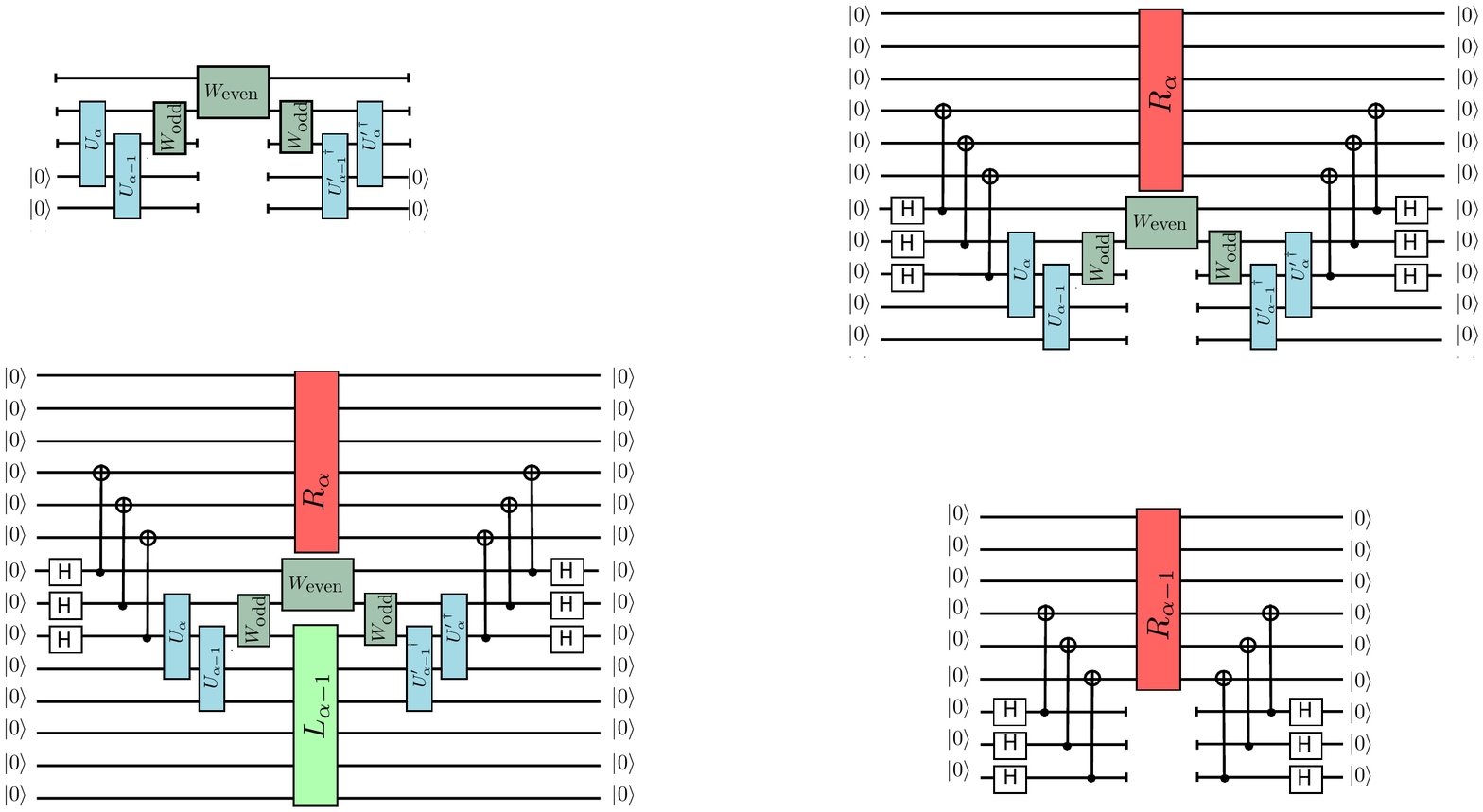}
 \nonumber\\
&=&\includegraphics[scale=0.6,valign=c]{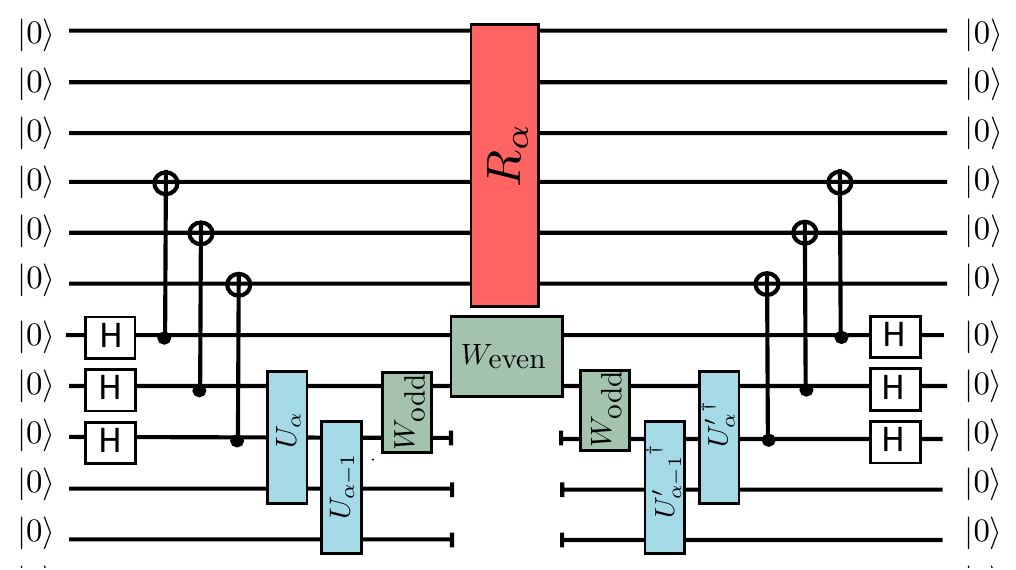},
\label{eq:RmixedIterative}
\end{eqnarray}
and 
\begin{eqnarray}
& &
\includegraphics[scale=0.6,valign=c]{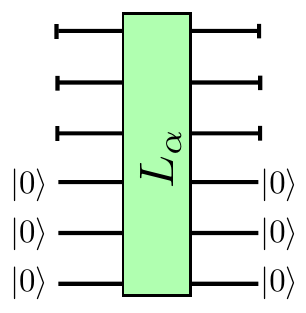}
=\includegraphics[scale=0.6,valign=c]{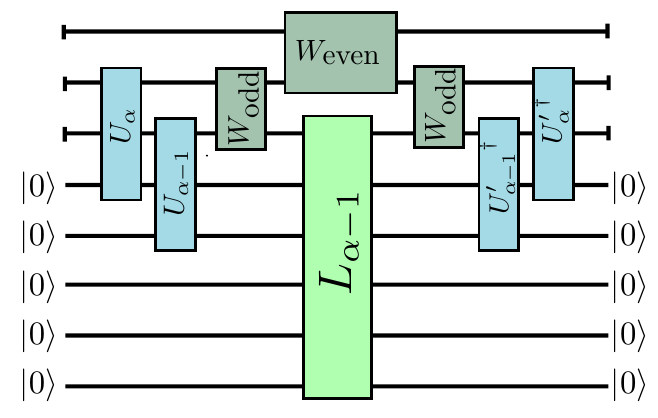}.
\label{eq:LmixedIterative}
\end{eqnarray}
When \SEq(\ref{eq:EvolutionOverlapDivided}) is optimised alongside \SEqs(\ref{eq:RmixedIterative}) and (\ref{eq:LmixedIterative}), the resulting evolution is equivalent to the TDVP equations over the variational parameters of $U$. Indeed, as noted in the Methods, setting the derivative of \SEq(\ref{eq:EvolutionOverlapDivided}) with respect to $U'$ equal to zero recovers the usual TDVP equations. the present implementation benefits from the encoding of the tangent space to the variational manifold in addition to the state in the unitaries $U_\alpha$. 

\vspace{0.1in}
\noindent
{\it For translationally invariant systems}, the environment unitaries $R_\alpha$ and $L_\alpha$ become independent of site and the iterative updates in \SEqs(\ref{eq:RmixedIterative}) and (\ref{eq:LmixedIterative}) reduce to the fixed point equations giving the highest weight right and left eigenvectors of the transfer matrix
\begin{eqnarray}
E^{U}_{U'}&=&\includegraphics[scale=0.8,valign=c]{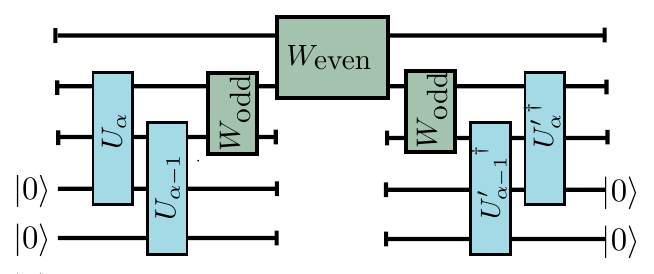}.
\label{eq:MixedTransferMatrix}
\end{eqnarray}
At this point we make a further simplification to allow for easier implementation on NISQ devices. We reduce to a first order Trotterization scheme take account of only $\exp[ i H_{\hbox{odd}} dt ]$. Taking advantage of the translational invariance or $U$ and $U'$ appears to reduce these errors somewhat. With this simplification the circuits in \SEqs(\ref{eq:EvolutionOverlapDivided}), (\ref{eq:RmixedIterative}) and (\ref{eq:LmixedIterative}) reduce to 
%
\begin{eqnarray}
& &
 \langle \psi (X(t+dt))| e^{i H dt} |\psi (X(t))\rangle
 \nonumber\\
&=&\includegraphics[scale=0.6,valign=c]{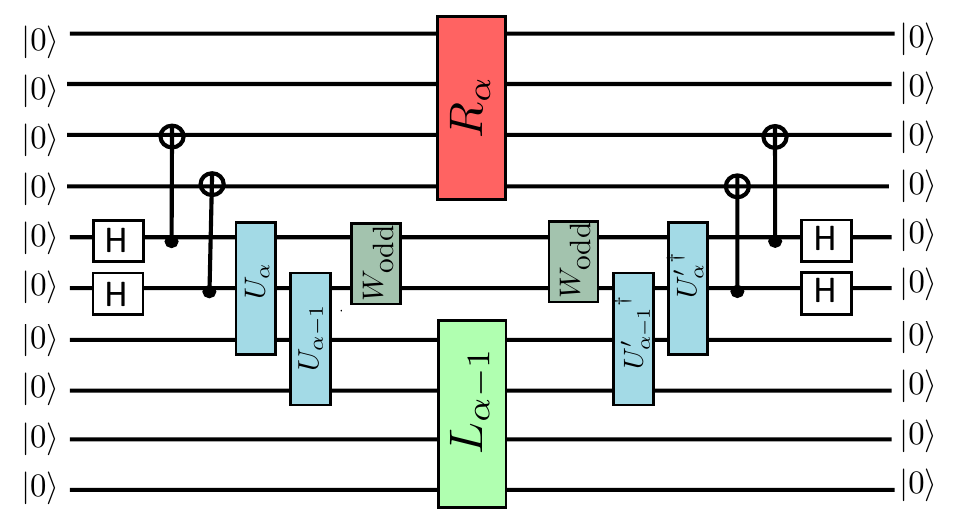},
\label{eq:EvolutionOverlapDividedSimpler}
\end{eqnarray}
for the overlap, and 
\begin{eqnarray}
& &
\includegraphics[scale=0.6,valign=c]{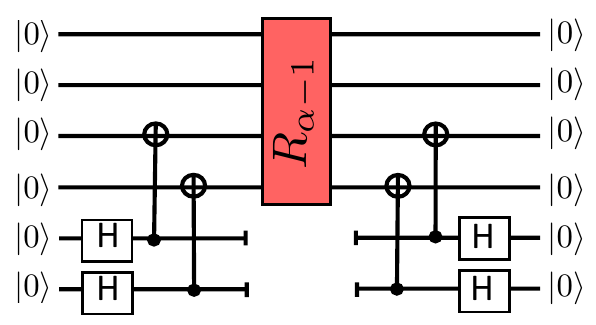}
 \nonumber\\
&=&\includegraphics[scale=0.6,valign=c]{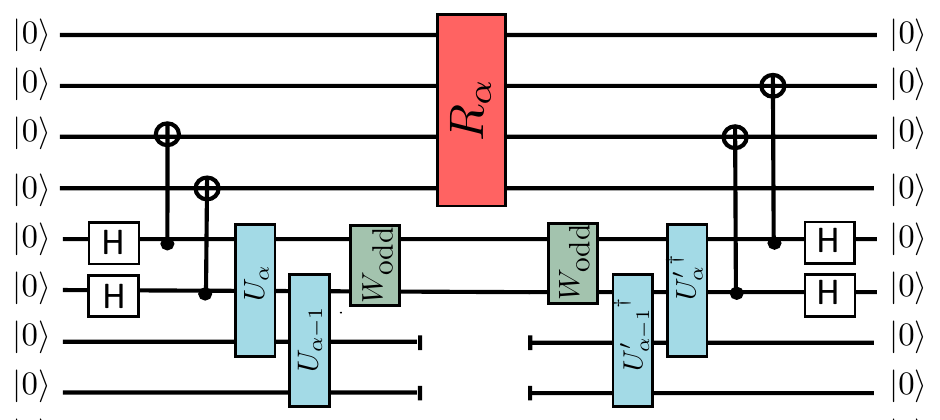},
\label{eq:RmixedIterativeSimpler}
\end{eqnarray}
and 
\begin{eqnarray}
& &
\includegraphics[scale=0.6,valign=c]{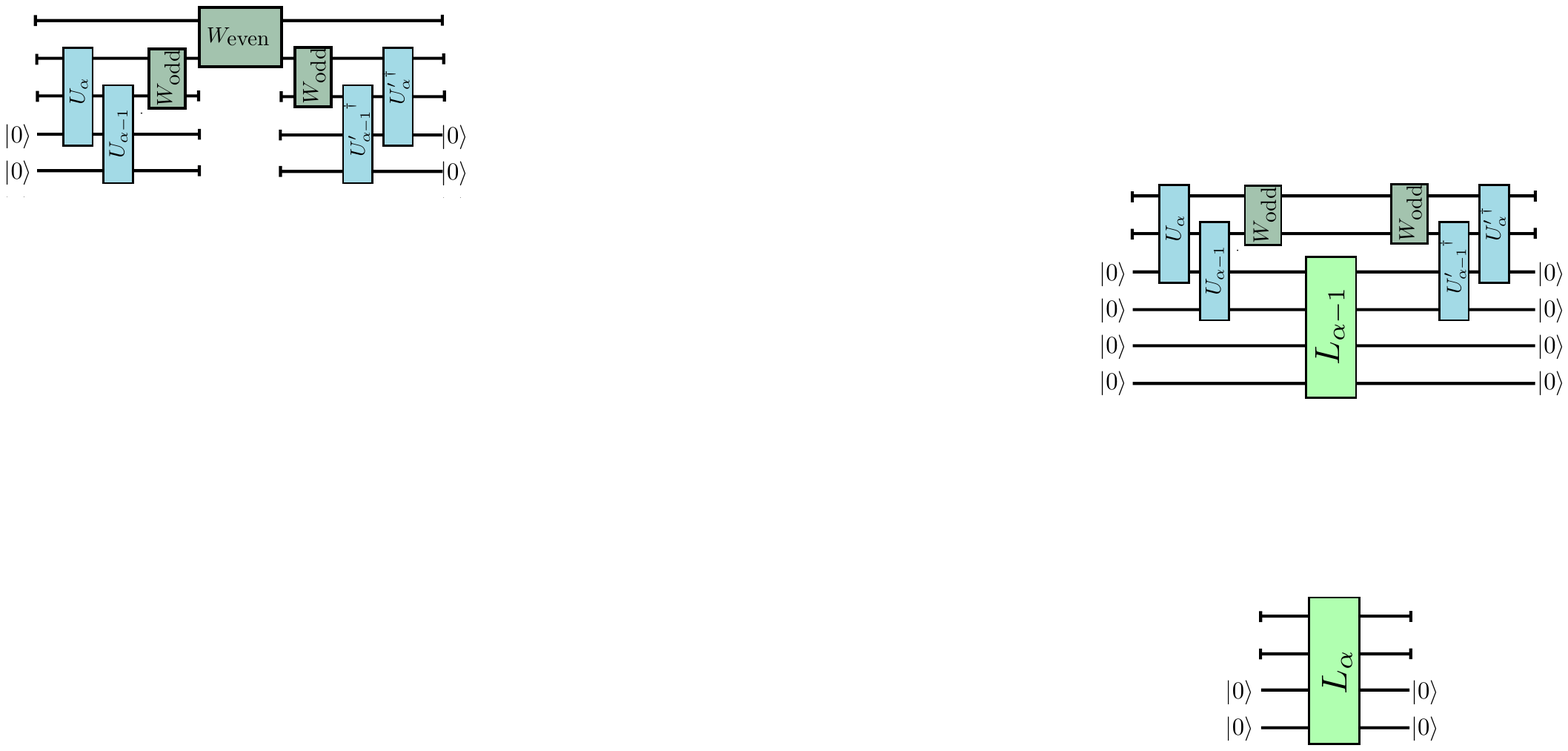}
=\includegraphics[scale=0.6,valign=c]{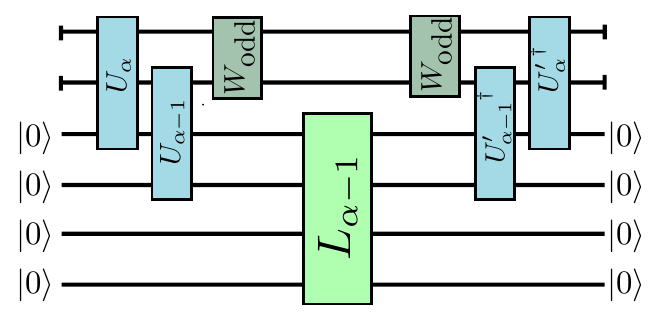}
\label{eq:LmixedIterativeSimpler}
\end{eqnarray}
for the fixed-point equations.

The simulations presented in the main body of the paper are carried out for $D=2$. 
This is mainly to obtain compact circuits that are feasible to implement on available (or imminently available) NISQ machines. Our code (in both Cirq and Qiskit) can run with arbitrary bond order. Restricting to bond order $D=2$ recovers the circuits shown in Fig. 3 in the main text.

\subsection{Implementing Quantum Circuit MPS}

\subsubsection{Finding the Environment}
\begin{figure}[htbp]
    \includegraphics[width=1\linewidth]{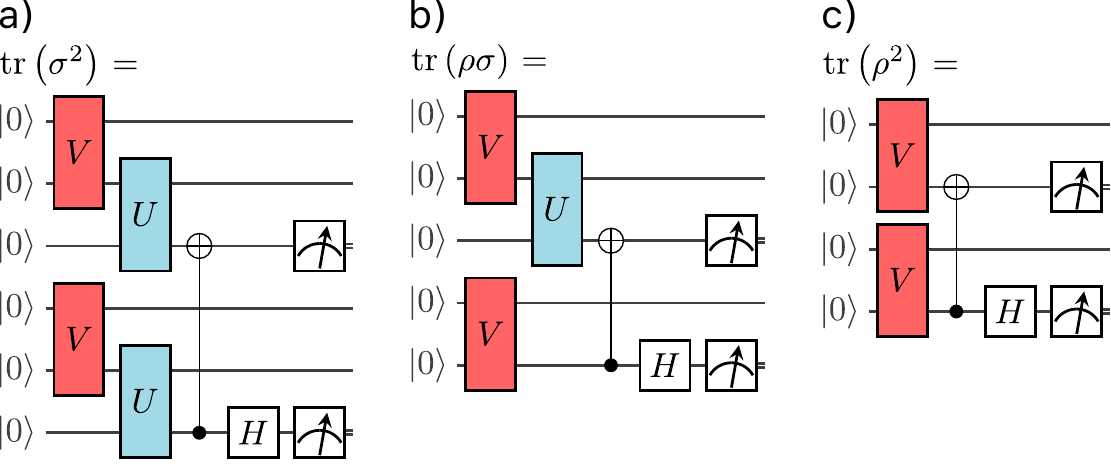}
\caption{{\bf Quantum circuits to find the environment:} In order to determine an approximation to the overlap, the fraction of $00$ outputs is measured on the measurement symbols.}
\label{fig:environment}
\end{figure}

In order the determine the environment on the quantum chip we need to implement the equality of Fig.~2 from the main paper.
Supplementary Fig.~\ref{fig:environment} details the quantum circuits required to do so. 
We use the SWAP test \cite{swap_test} to compare both sides of the equation. The minimal application of this  is illustrated by the circuit shown in Supplementary Fig.~\ref{fig:environment}b. 
However, the overlap $F(\rho, \sigma) = \tr(\rho \sigma)$ is not necessarily maximised by $\rho = \sigma$. In fact, $F(\rho, \rho)=\tr(\rho^2)$ is the the purity, so that if the reduced density matrices of Supplementary Fig.~\ref{fig:environment}b are mixed. The optimizer will incorrectly try to increase the purity. We circumvent this by using a quantity related to the  trace distance 
$ \tr((\rho-\sigma)^\dagger(\rho-\sigma))$, which is minimized at $\rho=\sigma$.
The additional circuits required to determine the trace distance are shown in Supplementary Figs.~\ref{fig:environment}a and  c.

\subsubsection{Finding the Overlap}

In order to time evolve a matrix product state on a quantum computer, we minimize the overlap between the perfect evolution of the state from time $t$ to $t+\delta t$ and the manifold of matrix product states of the available bond dimension. In the infinite translationally invariant case, the definition of the overlap requires care. 
It can be given as the largest eigenvalue of the {\it mixed transfer matrix}.
 

\vspace{0.1in}
\noindent
{\it The overlap as largest eigenvalue:}
 The overlap is the result of applying the matrix of \SEq(\ref{eq:MixedTransferMatrix}) an infinite number of times to boundary vectors. 
 The result (with appropriate normalisation) is equal to the largest eigenvalue. In order to find the largest eigenvalue, we use a variational method to find the largest eigenvalue-eigenvector pair. Since the matrix in \SEq(\ref{eq:MixedTransferMatrix})  is not Hermitian, we cannot use the Rayleigh-Ritz variational principle. 
Instead, we minimize the distance between the result of applying the transfer matrix to a vector, and the product of that vector with a candidate eigenvalue. Denoting the matrix of \SEq~\ref{eq:MixedTransferMatrix} as $E^{U}_{U'}$, we seek to solve:
\begin{equation}
            \max_{\eta}\min_{r}{\norm{E^{U}_{U'} r-\eta r}^2}.
\end{equation}
In the applications considered in this work, we need only the overlap between very similar states, and can take advantage of the fact that we know that $\eta$ can deviate from $1$ only by a correction of order $\delta t$. As a result, we have found it effective to apply a heuristic, wherein we minimise over both $\eta$ and $r$, using an appropriate initial condition for $\eta\sim 1$, and restart the optimization if it fails to attain $\eta\sim1$ at its conclusion: 
\begin{equation}
    \min_{\eta, r}{\norm{E^{U}_{U'} r-\eta r}^2 = \min_{\eta, r} v(\eta, r)},
\label{eq:minimisation}
 \end{equation}
where we have defined our objective function $v(\eta, r)$. In practice, only rarely are repetitions required to find the largest eigenvalue and even then a small number of repetitions suffices (See Supplementary Fig.~\ref{fig:Envperformance}).
  
\vspace{0.1in}
 \noindent
{\it Solving \SEq~(\ref{eq:minimisation}) on a quantum circuit:}
Expanding \SEq~(\ref{eq:minimisation}) allows us to put it in a form that can be implemented as quantum circuits:
\begin{equation}
            v(\eta, r) = r^\dagger{E_{U'}^U}^\dagger E_{U'}^U r+\abs{\eta}^2r^\dagger r- r^\dagger E_{U'}^U r-r^\dagger {E_{U'}^U}^\dagger r.
 \label{eq:minimisationexpanded}           
\end{equation}
\begin{figure}[b]
            \includegraphics[width=\linewidth]{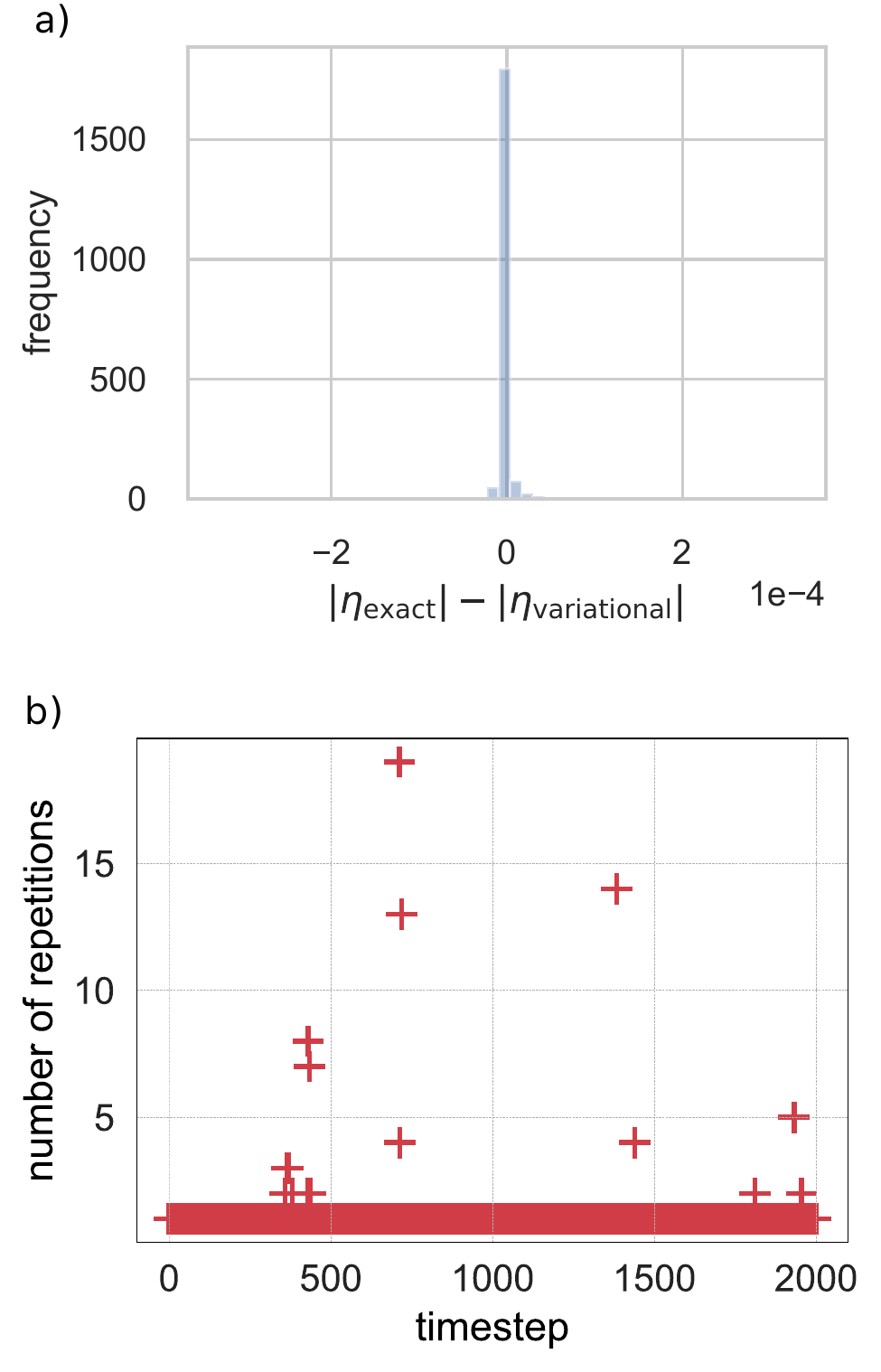}\\
\caption{{\bf Performance of finding the right environment with \SEq~(\ref{eq:obj})}
a) Distribution of results of minimizing \SEq~(\ref{eq:obj}:) a comparison of the exact and variational largest eigenvalue of the mixed transfer matrices for $2000$ steps along a TDVP trajectory, with $\delta t=0.01$, using the BFGS optimizer provided in scipy. 
b) Number of repetitions vs. time:  Number of repetitions required at each timestep in (a).}
  \label{fig:Envperformance}
\end{figure}
The first two terms can be implemented on quantum circuits as shown in Figs.~\ref{fig:mixed_environment}d and ~\ref{fig:mixed_environment}f ( a and b for the corresponding left eigenvectors). 
The objective function is found by measuring in the computational basis, determining the probability of the bit string of all $0$s, and taking the square root of the corresponding probability (since we know the result should be real, there is no problem determining the phase).
The same is not true of the third and fourth terms of \SEq~(\ref{eq:minimisationexpanded}), for which there is an undetermined phase.
At the minimum, the sum of the last two terms will be equal to the largest eigenvalue, which is real to $O(dt^2)$, since the largest eigenvalue of $E_U^U$ is $1$. 
 Therefore, if we minimize the associated objective function:
 \begin{equation}
            v'(\eta, r) = r^\dagger{E_{U'}^U}^\dagger E_{U'}^U r+\abs{\eta}^2r^\dagger r- 2\abs{r^\dagger E_{U'}^U r}, \label{eq:obj}
 \end{equation}
    by taking the square root of the corresponding probabilities (Supplementary Fig.~\ref{fig:mixed_environment} b and e),
    the $\mathrm{argmin}$ will coincide with that of \SEq~(\ref{eq:minimisation}) (to $O(dt^2)$), and we can determine the eigenvalue and eigenvector.
    In practise, we often achieve an accuracy much greater than $O(dt^2)$. 
    Supplementary Fig.~(\ref{fig:Envperformance}) gives data demonstrating the accuracy and reliability of this method of determining the overlap.

\begin{figure*}[ht]
        \centering
        \includegraphics[width=\linewidth]{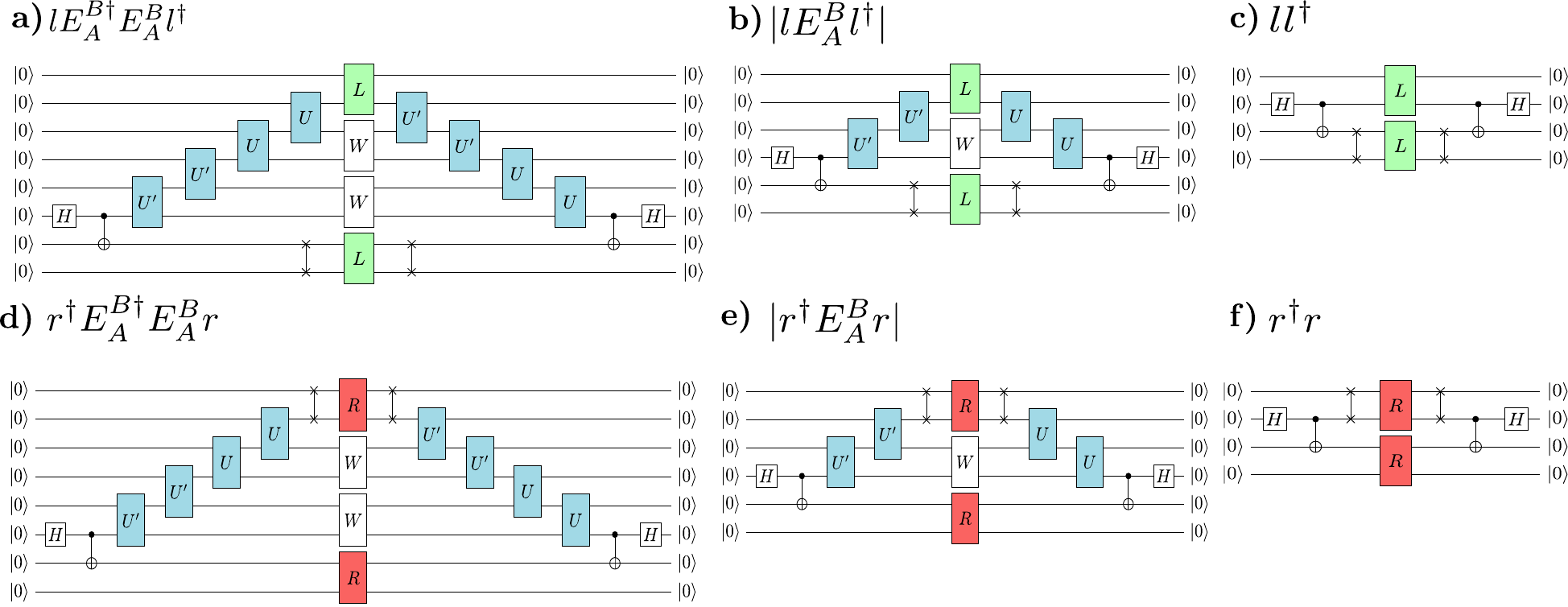}
        \caption{{\bf Diagrams required to determine the mixed environment:} Panels a), b) and c) show the circuits used to construct the left mixed environment and panels d), e) and f) those used to construct the right mixed environment. The probability of measuring the all 0s bit string is measured then the real square root is taken to calculate the labelled expression.}\label{fig:mixed_environment}
\end{figure*}

\subsubsection{Ans\"atze}\label{sec:app_ansatze}
The implementation of iMPS algorithms on quantum computers require factorisations of the large unitaries in which the tensors are embedded. Different ans\"atze make different tradeoffs between gate depth, gate availability on typical QC implementations, and expressibility. The different unitaries ($U$ and $V$, containing the state tensor and the environment tensor respectively), can be factorised in different ways, to take into account their different structures.  At low bond dimension, full parametrisations of the important matrix spaces are possible. 

\vspace{0.1in}
\noindent
{\it The state unitary $U$:}
 For this work, we used the generic ansatz for $U$ reported in Fig. 4a  in the main body of the paper.  
 We found it to be a workable tradeoff between expressibility and depth in the generic case.
 However, ground state problems often have deeper structure in the form of symmetries, and that structure can be exploited for more efficient circuit ans\"atze, and more effective optimization. 

 As an example, the transverse field Ising model used as an example in the main text exhibits time reversal invariance, parity symmetry, and has a conserved quantity (the total magnetization).
 While in Fig.~4b in the main text we use the general ans\"atz of Fig.~4a,

 These properties constrain the form of the true ground state of the system, and its MPS approximation. 
 The transverse field Ising model Hamiltonian is real, and thus the model exhibits time reversal invariance. 
 The MPS form of the ground state should be made up of real tensors~\cite{pollmann_symmetry}.
 In Supplementary Fig.~\ref{fig:tri_symmetry}, we demonstrate the resulting speedups for two different real ans\"atze.
 Note in particular that for the ansatz of Supplementary Fig.~\ref{fig:tri_symmetry}, the optimization produces optimal results (i.e. equivalent to D=2 classical), for a state tensor of only depth 4. 
 \begin{figure}[ht]
     \includegraphics[width=\linewidth]{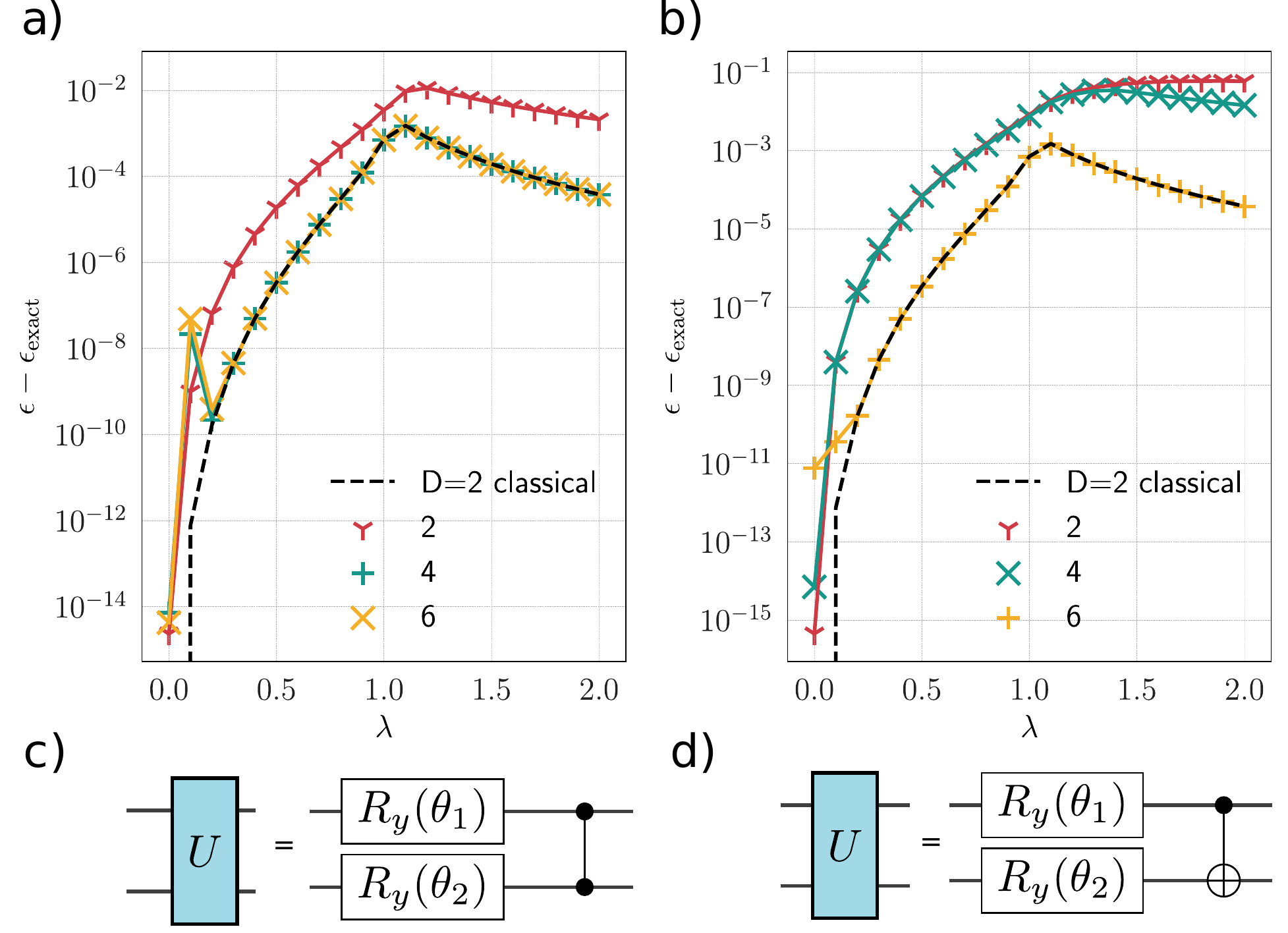}
 \caption{{\bf Time Reversal Symmetry in optimization of the transverse-field Ising model: } a) and b) Deviation from analytical ground state energy density of the transverse-field Ising model as a function of $\lambda$. Dashed line is the result of classical iDMRG at D=2, other lines/markers are the result of quantum variational optimization at different ansatz depths, given in the legend. c) and d) Ans\"atze used to produce figures a) and b), respectively.}\label{fig:tri_symmetry}
 \end{figure}

\vspace{0.1in}
\noindent
{\it Representing the environment:}
The simulations presented in the main paper were performed at $D=2$.  At $D=2$, the depth required for a full parametrisation of the environment is easily accessible, and a minimal parametrisation is available. The resulting circuits allow us to calculate important quantities (the Schmidt coefficients, for example) classically, and point the way to natural generalisations for $D>2$.

\vspace{0.1in}
\noindent
{\it A full parametrisation of the right environment at $D=2$:}
The environment -- the right fixed point, $r$, of the MPS transfer matrix~\cite{orus_review} -- is an Hermitian, positive definite matrix, with trace 1.
$V$ embeds the Cholesky decomposition of the right environment in a unitary.
To obtain the right environment, we parametrise $V$ as shown in Supplementary Fig.~\ref{fig:env_par_acr}.

Since the environment $r$ has trace $1$, the Cholesky factor $r^{\frac{1}{2}}$ has Frobenius norm 1.
First consider the matrix in Supplementary Fig.~\ref{fig:env_par_acr}a.
This spans all possible diagonal environment square roots, since $\tr(\Lambda^\dagger \Lambda)=1$, $\Lambda^\dagger \Lambda \geq 0$, and $\cos(\gamma)$ is surjective onto $[-1, 1]$.

\noindent
This parametrisation has some advantages: \\
\noindent
i. It is minimal, depending  upon only 3 real parameters; \\
\noindent
ii. The eigenvalues of the environment are classically accessible. The purity, Von Neumann entanglement entropy {\it etc}. are all accessible with simple calculations if we know the parameters of the ansatz. 

\begin{figure}[ht]
    \includegraphics[width=\linewidth]{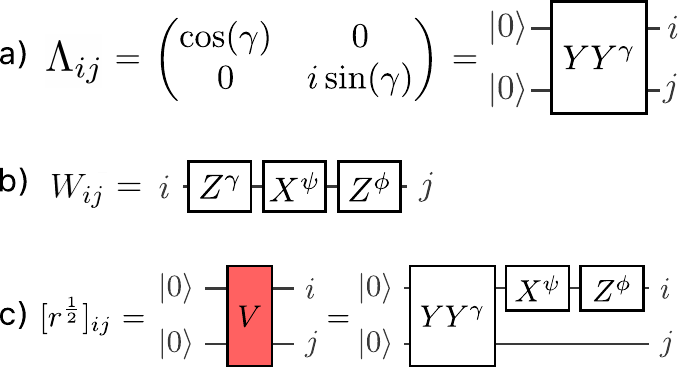}
            \caption{{\bf Parametrisation of the environment $V$, used throughout:} a) shallow quantum circuit to represent the positive diagonal matrix used in Supplementary Eqs.  (\ref{eq:UforR})  and (\ref{eq:DiagonalV}). b) The right environment can be constructed from this using Supplementary Eq. (\ref{eq:UforR}) in which $W$ is represented on the circuit as shown. c) The unitary $V$ is constructed as shown. }
            \label{fig:env_par_acr}
\end{figure}

\vspace{0.1in}
\noindent
{\it Parametrising the right environment at} $D>2$:
It is not obvious how best to parametrise the right environment for higher bond dimension, for which we will not be able to efficiently access the full space of matrices. 
We have, however, found a natural generalisation of Supplementary Fig.~\ref{fig:env_par_acr} to be surprisingly effective in practise.
The environment at any bond dimension can be unitarily diagonalised, with real, positive eigenvalues:
\begin{equation}
	r = W\Lambda^2 W^\dagger.
	\label{eq:UforR}
\end{equation}
We can assume that the unitary $W$ can be absorbed by changing the parameters of the state tensor unitary $U$. 
This amounts to diagonalising the environment, an important step in the development of classical MPS algorithms \cite{mps_representations}.
The task that remains is to express the positive diagonal matrix $\Lambda$ within a shallow quantum circuit, by embedding it across the legs of a unitary $V$:
\begin{equation}
    (V\ket{0})_{ij} = \Lambda_{ij}.
    \label{eq:DiagonalV}
\end{equation}
where the indices on the vector $V\ket{0}$ should be understood as denoting the elements of that vector when reshaped into a matrix. 

For bond dimension $2$ we have made use of the unitary $V = YY^\gamma$ to express this diagonal. 
At higher bond dimensions, $V$ generates a map from qubits $(1, \ldots, n) \rightarrow (n+1, \ldots, 2n)$, where $D = 2^n$. 
We can apply the ansatz of Supplementary Fig.~\ref{fig:env_par_acr}a to the all pairs of qubits $(i, n+i), i \in [1, ... n]$, each with a different parameter $\gamma_k$.
In doing so we will parametrise a subset of the possible diagonal matrices, of the form $\Lambda(\vec{\gamma}) = \otimes_{i=1}^n \Lambda(\gamma_k)$, where $\Lambda_{ij}$ is the matrix of Supplementary Fig.~\ref{fig:env_par_acr}a.
Whilst this parametrisation no longer explores the full space of diagonal matrices, certain properties -- the entanglement entropy, for example -- can still be efficiently calculated from the parameters of the ansatz.

We have found such a constant depth ansatz for $r$ to be effective for the problem of ground state finding, for all bond dimensions for which the algorithm can be effectively classically simulated.
More work is required to produce theoretical guarantees that this should be the case in general.

\vspace{0.1in}
\noindent
{\it Alternative representations of the environment:}
 When considering the quantum circuits used to implement the time-dependent variational principle, it is convenient to use a different representation of the environment as indicated in Supplementary Fig.~\ref{fig:alt_param}a. This restructuring allows us to trade depth for qubits, and makes manifest the time reversal symmetry of the TDVP circuits. It also allows us to explore non Hermitian tensors variationally. This is crucial in the calculation of the overlap.
The parametrisation of the tensor $R$ (Supplementary Fig.~\ref{fig:alt_param}b) retains the benefits of the parametrisation of Supplementary Fig.~\ref{fig:env_par_acr}c, in particular, the eigenvalues of the environment (the square of the Schmidt coefficients across each bond) are: $\lambda_1 = \cos^2(\theta), \lambda_2 = \sin^2(\theta)$.
    \begin{figure}
        \centering
        \includegraphics[width=\linewidth]{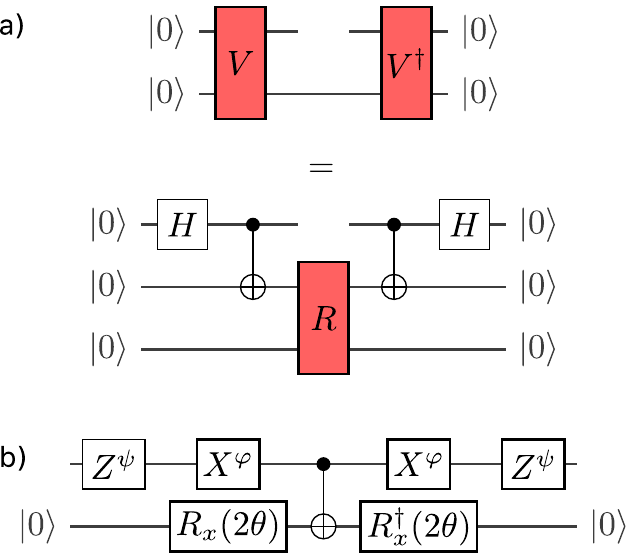}
        \caption{{\bf Alternative representation of the environment and its circuit realisation:} a) The alternative representation of the environment, used in the TDVP circuits and b) the corresponding full parametrisation of the unitary $R$.}
        \label{fig:alt_param}
    \end{figure}
    \begin{figure}
        \centering
    \end{figure}

\subsubsection{Optimization Methods}
Variational algorithms require the use of effective optimization algorithms.
In the main text, we have made use of several - the standard algorithms provided in scipy, and a coordinate optimization method named Rotosolve. The details of this algorithm --- and its name --- were presented in Ref.~[\onlinecite{Ostaszewski2021structure}]. The same algorithm having been developed independently by several authors time\cite{vidal2018calculus,parrish2019jacobi,nakanishi2020sequential}
In this section we detail a necessary modification to the Rotosolve algorithm, and compare the performance of each on our setup.

\vspace{0.1in}
\noindent
{\it Doubled Rotosolve:}
Rotosolve is an optimization method for quantum circuits.
As detailed in Ref.~\cite{Ostaszewski2021structure}, it takes advantage of the fact that for quantum circuits $U(\vec{\theta})\ket{0}$, in which each $\theta_k$ parametrises a single exponential of a Pauli string ($e^{-i\theta_k S}$, $S^2=I$), the expectation values of local observables must vary sinusoidally as a function of $\theta_k$. 
The details of the resulting sinusoid can be determined by three measurements at different values of $\theta_k$, and once determined, the minimum for that parameter (for the current values of the rest of $\vec{\theta}$) can be attained immediately.
Analytic gradient methods like Rotosolve can lead to dramatic ($\sim100\times$) reduction in the number of measurements required to optimize a circuit. 

Rotosolve as introduced in Ref.~\cite{Ostaszewski2021structure} requires that each parameter control at most one gate. For circuits such as those shown in Figs.~\ref{fig:environment}a-c), where each parameter controls more than one gate,  a small modification to the Rotosolve algorithm is necessary (see also Ref.[\onlinecite{nakanishi2020sequential}]).

\vspace{0.1in}
\noindent
{\it Doubly Sinusoidal Expectation Values:}
A derivation of the original Rotosolve algorithm is given in the appendix of Ref.~\cite{Ostaszewski2021structure}.
Here we present a brief extension of this derivation to the situation where a given parameter controls two gates.
Consider a generic quantum circuit, in which each element is the parametrised exponential $U_i = e^{-i\theta_i S_i}$ of some Pauli string $S_i$, $i \in [1\dots N]$, $S_i^2=I$.
Consider the expectation value of an operator $\hat{O}$ in the output of this circuit, as a function of one of the $\theta_i$, in a circuit consisting of $N$ gates depending on $M<N$ parameters, where at most two identical gates $U_k$ depend upon the same parameter $\theta_k$.
Using the expansion $U_k(\theta_k) = e^{-i \theta_k S_k} = \cos(\theta_k/2)\mathbb{I} + \sin(\theta_k/2) S_k$, and defining the quantities:
\begin{align}
    A &= \ev{\hat{O}}_{\theta_k=0}+\ev{\hat{O}}_{\theta_k=\pi}, \\
    B &= \ev{\hat{O}}_{\theta_k=0}-\ev{\hat{O}}_{\theta_k=\pi}, \\
    C &= \ev{\hat{O}}_{\theta_k=\pi/2}+\ev{\hat{O}}_{\theta_k=-\pi/2}, \\
    D &= \ev{\hat{O}}_{\theta_k=\pi/2}-\ev{\hat{O}}_{\theta_k=-\pi/2}, \\
    E &= \ev{\hat{O}}_{\theta_k=\pi/4}-\ev{\hat{O}}_{\theta_k=-\pi/4}, 
\end{align}
and the combinations:
\begin{equation*}
    a = \frac{1}{4}\left(2E-\sqrt{2}D\right), b = \frac{1}{4}(A-C), c =\frac{1}{2}D, d = \frac{1}{2} B,
\end{equation*}
one can show that:
    \begin{align}
        \ev{\hat{O}}_{\theta_k} &= a\sin(2\theta_k) + b\cos(2\theta_k) +c\sin(\theta_k) + d\cos(\theta_k) \nonumber\\
                                &= P\sin(2\theta_k+\phi) + Q\sin(\theta_k +\psi)\label{eq:rotosolve}
    \end{align}
where $P = \sqrt{a^2+b^2},~\phi = \arctan_2(a, b),~Q = \sqrt{c^2+d^2},~\psi = \arctan_2(c, d)$.
Unlike the original Rotosolve algorithm, \SEq~(\ref{eq:rotosolve}) is not a single sinusoid but a sum of sinusoids with doubled frequencies and different amplitudes and phases. 
It can no longer be minimized analytically. 
However, the spirit of the Rotosolve algorithm remains, and one proceeds as follows:
\begin{enumerate}
    \item Perform the 6 measurements required to determine $A, B, C, D, E$.
    \item Perform a classical scalar minimization on the function $\langle \hat{O} \rangle_{\theta_k}$ (whose coefficients have now been determined). This can be done very quickly on the fly, or precomputed and interpolated for arbitrary $P, Q, \phi, \psi$. 
    \item Set the variable $\theta_k$ to its corresponding minimum.
\end{enumerate}
The resulting algorithm suffers no significant loss of performance over Rotosolve, and only requires 3 more measurements per iteration.

 \begin{figure}
     \centering
     \includegraphics[width=0.8\linewidth]{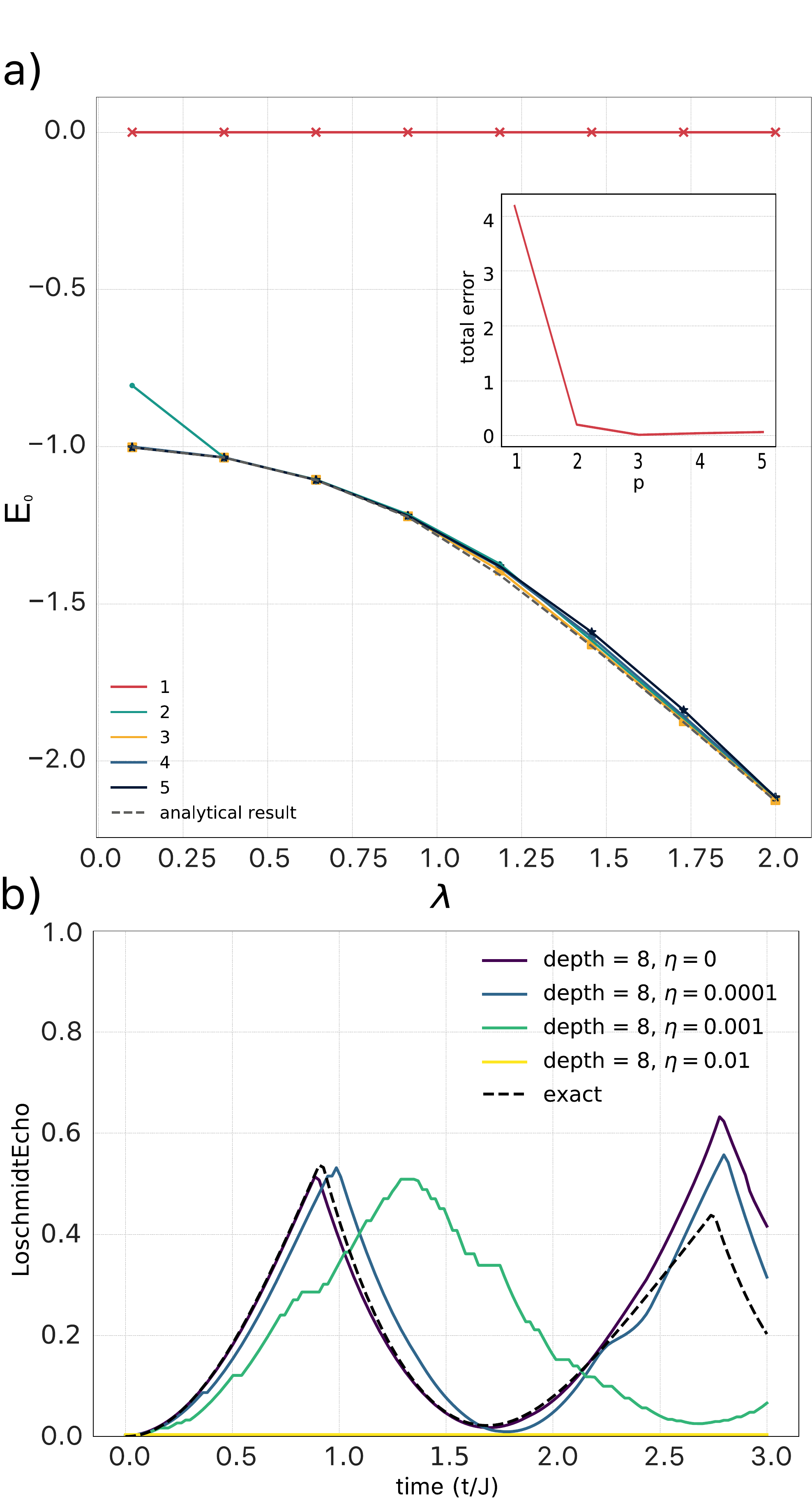}
\caption{{\bf Effect of finite gate fidelity upon optimisation and evolution:}  a) Effect of noise upon time evolution, $\eta$ is depolarising probability. Increasing noise deteriorates but does not destroy the appearance of the dynamical phase transition peaks. 
         b) Optimization, for depolarising probability $\eta=1\times10^{-3}$, and varying circuit depth (legend). Inset shows total deviation from exact curve as a function of depth ($p$). Note the tradeoff between circuit depth and sensitivity to noise. 
         For computational expediency, expectation values are calculated exactly from noisy circuits.  }
\label{supfig:noise}
\end{figure}

\vspace{0.1in}
\section{Supplementary Discussion}

In the following, we provide some further technical details of our results 


 
 \vspace{0.1in}
 \noindent
 {\it Finite Gate Fidelity and Restricting Circuit Depth:}
 Finite gate-fidelity implies a maximum reliable depth of quantum circuit. As evident from Figs. 2 and 3,  passing from the simplest task of representing a quantum state through the more complicated tasks of optimising and  time-evolving states, the required circuits are broader and of greater depth. Time-evolution is therefore much more susceptible to gate errors than simply representing a state. Whilst it might be possible to represent and measure properties of a rather high bond-order state, time-evolution is inevitably restricted to lower bond order. 

 Supplementary Fig.~\ref{supfig:noise}a shows how the attainable groundstate fidelity of the transverse-field Ising model ${\cal H} = \sum_i{\sigma_{i}^z \sigma_{i+1}^z+\lambda \sigma_i^x}$ changes with $\lambda$ for fixed noise strength $\eta=1\times10^{-3}$. 
The same gate fidelity has a much more severe effect upon simulations of dynamics. Supplementary Fig.~\ref{supfig:noise}b shows how a simulation at $D=2$ is degraded by this error rate. 

This formalism allows us to make a variety of tradeoffs between noise and performance, depending upon the resources available on the quantum computer. One can increase the expressibility of the circuit by increasing the depth of the ansatz.
It is also possible to modify the variational space by using a shallow ansatz at a larger bond order, which requires a greater qubit count. 
It remains to be seen how each of these options will perform on a near term chip. 


 \vspace{0.1in}
 \noindent
 {\it Construction of the Poincare Map:}
To produce a Poincar\'e map, the parameters that define the quantum state are recorded as they evolve over time. When a chosen parameter crosses a particular value in the positive direction the values of the other parameters are plotted. The values of these crossings indicated in Fig. 5c of the main text are obtained by polynomial interpolation of the values at discretised time intervals obtained from the quantum circuit. This proceeds as follows: Around the approximate crossing time, polynomial function interpolation is used to estimate the evolution of all parameters. A root finding algorithm is used on the  interpolated polynomial function of the chosen parameter to get an estimated time of crossing. Finally this time is input into the other interpolated functions to get estimates of all the parameters at the crossing time.

\bibliographystyle{naturemag}
\bibliography{BibliographySubmission.bib}